\documentclass[12pt]{article}
\usepackage[utf8]{inputenc}
\usepackage{amsmath}
\usepackage{amsthm}
\usepackage{enumitem}
\usepackage{multirow}
\usepackage{amsfonts}
\usepackage{amssymb}
\usepackage{tabularx}
\usepackage{graphicx}
\usepackage{subcaption}
\usepackage{placeins}
\usepackage{authblk}
\usepackage{afterpage}
\usepackage{setspace}
\usepackage{mathtools}
\usepackage[toc,page]{appendix}
\usepackage{pdfpages}
\usepackage{listings}
\usepackage{bm}
\usepackage{mathrsfs}
\usepackage{natbib}
\usepackage{algorithm}
\usepackage[algo2e]{algorithm2e}
\usepackage{url}
\usepackage{mdframed}
\usepackage{soul}
\usepackage{array} % for specifying fixed width in table
\newcommand{\pkg}[1]{{\normalfont\fontseries{b}\selectfont #1}}
\newtheorem{theorem}{Theorem}
\newtheorem{lemma}[theorem]{Proposition}

\newcommand{\blind}{0}

\begin{document}

\def\spacingset#1{\renewcommand{\baselinestretch}%
{#1}\small\normalsize} \spacingset{1}

\if0\blind
{
  \title{\bf Modeling the Marked Presence-only Data: A Case Study of Estimating the Female Sex Worker Size in Malawi}
  \author{Ian Laga, Xiaoyue Niu, and Le Bao \thanks{This work was supported by the National Institute of Allergy and Infectious Diseases of the National Institutes of Health under award number R01AI136664. Correspondence to: lebao@psu.edu}\\
    Department of Statistics, Pennsylvania State University}
  \maketitle
} \fi

\if1\blind
{
  \bigskip
  \bigskip
  \bigskip
  \begin{center}
    {\LARGE\bf Modeling the Marked Presence-only Data: A Case Study of Estimating the Female Sex Worker Size in Malawi}
\end{center}
  \medskip
} \fi

\bigskip
\begin{abstract}
Certain subpopulations like female sex workers (FSW), men who have sex with men (MSM), and people who inject drugs (PWID) often have higher prevalence of HIV/AIDS and are difficult to map directly due to stigma, discrimination, and criminalization. Fine-scale mapping of those populations contributes to the progress towards reducing the inequalities and ending the AIDS epidemic. In 2016 and 2017, the PLACE surveys were conducted at 3,290 venues in 20 out of the total 28 districts in Malawi to estimate the FSW sizes. These venues represent a presence-only data set where, instead of knowing both where people live and do not live (presence-absence data), only information about visited locations is available. In this study, we develop a Bayesian model for presence-only data and utilize the PLACE data to estimate the FSW size and uncertainty interval at a $1.5 \times 1.5$-km resolution for all of Malawi. The estimates can also be aggregated to any desirable level (city/district/region) for implementing targeted HIV prevention and treatment programs in FSW communities, which have been successful in lowering the incidence of HIV and other sexually transmitted infections. 

\end{abstract}

\noindent%
\vfill

\newpage
\spacingset{1.5} % DON'T change the spacing!
\section{Introduction}
\label{sec:intro}

The work in this paper is motivated by the study of the HIV/AIDS epidemic, which is especially a public health threat in sub-Saharan African countries. Eastern and Southern Africa account for 45\% of HIV infections and 53\% of individuals living with HIV globally. The 90-90-90 treatment target set by UNAIDS aims to achieve three goals by the end of 2020: 90\% of all people living with HIV know their HIV status, 90\% of all people with diagnosed HIV infection will receive sustained antiretroviral therapy, and 90\% of all people receiving antiretroviral therapy will have viral suppression \citep{joint201490}. While saving lives has been very successful, the number of new HIV infections was not falling fast enough to meet the 2020 goal \citep{unaids2018data}. New HIV infections were reduced by only 18\% from 2010 to 2017, much lower than the 2020 target of 75\%. In order to decrease new HIV infections, it is first necessary to understand where those with AIDS live. A very large portion of new HIV infections occur in key populations like sex workers, men who have sex with men, and intravenous drug users. Female sex workers (FSW) are one of the largest groups and are 12 times more likely to have HIV than the general population \citep{unaids2014gap}. One study of 16 sub-Saharan African countries found the HIV prevalence for FSW to be above 37\% \citep{unaids2014gap}. It is clear that treating HIV positive individuals is not enough to meet the 90-90-90 goal, and success depends on reducing the spread of HIV in these key populations.

FSW populations are difficult to map directly because they represent a relatively small part of the population and have an incentive to hide their status. Small area estimates often suffer from large uncertainty due to small sample sizes. The University of North Carolina performed on-site surveying of venues believed to house FSW to estimate district-level FSW size in Malawi, which is by far the most detailed venue level data \citep{PLACEreport}. The venue is the unit of observation in the study. Their team published the Priorities for Local AIDS Control Efforts (PLACE) report, which documents how venues were discovered and visited and how FSW size estimates were calculated using probability weights. Only a subset of the reported venues were visited. Prior to surveying venues, the PLACE team obtained accurate counts of the total number of venues in each district. By mapping venues instead of FSW directly, the PLACE team can reach a much larger portion of the FSW population. 

In an ideal case, regression models to map population sizes are built on records of both where people live and where no people live. Records that include both pieces of information are known as ``presence-absence'' or ``used-unused'' data \citep{pearce2006modelling}. Regression methods can typically be applied directly to presence-absence data without much trouble. However, recording absences can be difficult, leading to presence-only data, which is data where only coordinates of observed sites are recorded, offering no information of presence for other locations. Most of the original presence-only literature used a ``pseudo-sampling'' approach, which samples background observations and assumes they are absences. Combining the pseudo-absences with the observed presences created a pseudo-presence-absence data set that could be modeled with more traditional techniques. We recommend \cite{pearce2006modelling} for a nice introduction. Maxent also emerged as a popular choice for modeling presence-only data \citep{phillips2004maximum, phillips2006maximum}.

Given the lack of information about absences, modeling abundance given presence-only data is challenging, so traditional models limit inference to binary presence-absence estimates, rather than any measure of cell-level abundance. To the best of their knowledge, \cite{pearce2006modelling} knew of no application modeling abundance given presence only. \cite{ward2009presence} later developed an EM approach which can estimate abundance given presence only when an estimate of population prevalence is available, but it can only handle binary presence-absence observations. Later researchers developed point process models for the data which overcame many of the existing problems. See \cite{banerjee2014hierarchical} and \cite{renner2015point} for some examples.

Additionally, the observations may be marked points. For marked point processes, the observations contain additional information about the observation. For example, the locations of trees constitute a point process, and the height or width of the trees are ``marks.'' \cite{banerjee2014hierarchical} offers a nice chapter on marked point processes.

%In this article, we aim to map FSW size at fine-scale cells by estimating the number of venues at the cell-level (venue count) and the FSW size at the venue-level (venue size). We estimate the number of FSW at the cell-level (FSW size) by taking the product of the estimated venue count and the estimated venue size for a given cell. Residual diagnostics verify the validity of the conditional independence between the venue count and the venue size. 

The PLACE data are an example of the marked presence-only data, where locations of visited venues constitute the point process, the visited venue sizes represent the marks, and the locations of unvisited venues are unknown. The PLACE report also provides the total number of venues in each district. In this article, we propose a Bayesian model-based approach that can map abundance given both presence-only data and the total number of observations. We utilize the PLACE data to estimate Malawi FSW sizes at fine-scale cells. Our proposed model is flexible and can account a wide range of data properties. Our model can handle cells with more than one observation, estimate abundance given the total number of observations, include random effects to reduce heterogeneity and account for systematic differences between areas, handle marked observations, incorporate uneven sampling efforts, and calculate credible intervals for domains that include multiple cells.
%The Bayesian framework allows for seamless credible interval calculations across multiple cells, in addition to the general flexibility of Bayesian models, such as incorporating predictor measurement error and prior information.
% \textcolor{red}{Very few of the existing methods can estimate abundance. The EM algorithm in \cite{ward2009presence} can estimate abundance, but cannot easily be extended to include random effects. Our proposed model discretizes the domain into cells and models the number of observed venues in each cell, approximating a point process model. While quadrat analysis is not new to spatial point patterns, we are unaware of its application to presence-only data. In our case, the presence-only data comes from a thinned point pattern, which represent the observed presences. We show that if the original cell-level counts come from a zero-inflated negative binomial distribution, then the observed venues also come from a zero-inflated negative binomial distribution, where only the mean is scaled by the sampling probability. The approximation both alleviates computational burden as well as makes it easier to incorporate the district-level venue counts. The Bayesian formulation is also desirable. First, informative priors based on previous studies can be included. Second, Bayesian computation allows us to easily account for measurement error in the predictors and include a calibration step to scale cell-level estimates such that the predicted number of venues in each district is equal to the known count.}

The paper is organized as follows. Section \ref{sec:Data} introduces the data and their special features and how to handle the spatial misalignment issue. In Section \ref{sec:NewMethod}, we introduce our choices of distributions and a calibrated Bayesian presence-only approach for modeling the FSW size at the cell-level. We validate our approach on a complete presence-absence subset of the PLACE data and via simulations in Section \ref{sec:validation}. We apply the method on the entire PLACE data in Section \ref{sec:RealData}, providing cell-level and district level estimates of FSW size, including credible intervals. Finally, Section \ref{sec:Discussion} covers concluding remarks and future work.

\section{Data and Special Features}
\label{sec:Data}
We combine multiple data sources in order to estimate the FSW size at a fine-scale grid. Our responses (FSW size at visited venues and the venue locations) come from PLACE, while the predictors come from a combination of environmental, demographic, and health data. Subsection \ref{sec:PLACE} further introduces the PLACE data. In subsection \ref{sec:aux}, we explore the auxiliary variables we use as predictors of FSW size. Data pre-processing and spatial alignment is handled in subsection \ref{sec:comb_data}.

\subsection{PLACE Data}
\label{sec:PLACE}

\begin{figure}[!tb]
    \centerline{\includegraphics[width=3in]{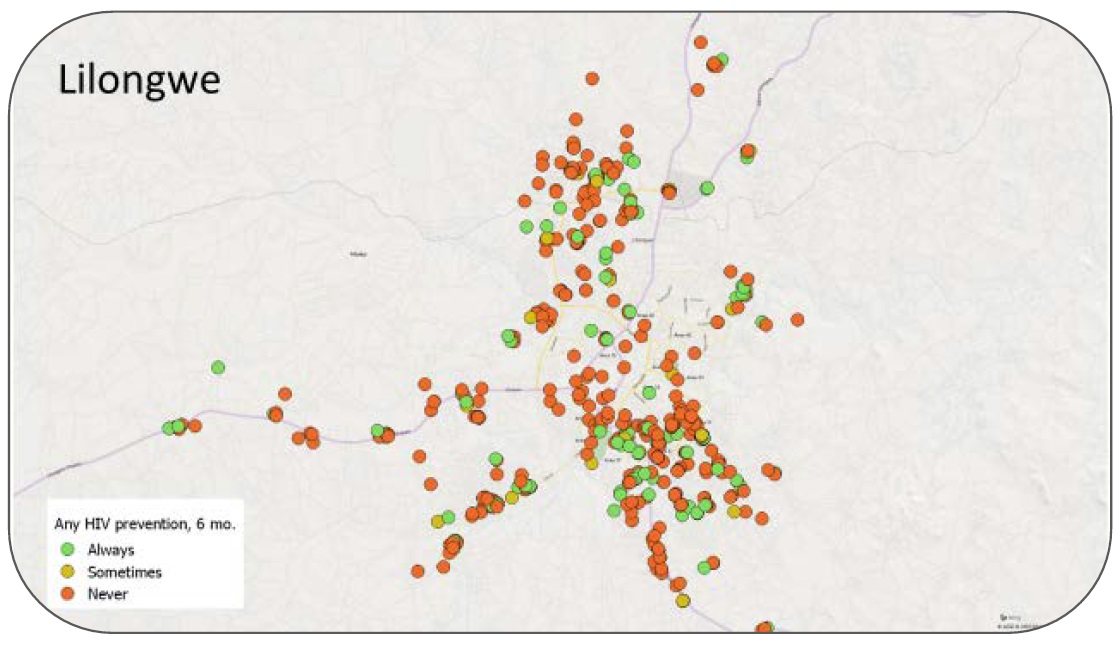}}
    \caption{An example of visited sites in Lilongwe, reprinted with permission from the PLACE report. Colors indicate whether any HIV prevention was available at the site in the last 6 months. HIV prevention activities include condom availability, HIV testing, and similar activities.}
    \label{fig:PLACE_sites}
\end{figure}

The Priorities for Local AIDS Control Efforts (PLACE) is funded by the United States Agency for International Development (USAID) and the United States President's Emergency Plan for AIDS Relief (PEPFAR) to understand the HIV epidemic and help reach the 90-90-90 target \citep{measuresource}. PLACE was developed to address the local behavior of HIV transmission. In 2016, The University of North Carolina implemented the PLACE I study in Malawi, surveying five districts (Lilongwe, Blantyre, Mangochi, Machinga, and Zomba) and one city (Mzuzu). Through additional funding, the PLACE II study was implemented in 15 additional districts, including the rest of Mzimba, the district for Mzuzu. The four main objectives enumerated in the PLACE report (2018) are: (1) to conduct programmatic mapping in selected district to identify venues where key populations can be reached, (2) to estimate the size of key population in each district who can be reached at venues, (3) to characterize HIV service coverage indicators for HIV programs reaching key populations, (4) in a subset of districts, to survey and test members of key population groups. Figure \ref{fig:PLACE_sites} includes a map of the visited venues in the Lilongwe district \citep{PLACEreport}. We can see that venues are highly clustered and typically lie along major roads or near the city center.

In order to achieve the first goal of PLACE, the team interviewed community informants to form a list of venues which were believed to host FSW, cleaned the list, and verified that the list was complete. The district-level venue counts were obtained for the 20 districts in PLACE I and II. The PLACE team visited a subset of venues. GPS locations were only recorded for these visited venues and thus formed the presence-only data. Interviews with knowledgeable personnel at each venue provided estimates of the number of FSW present at each venue. In PLACE I, the original plan was to visit all identified venues. However, due to time constraints, a convenience sample of venues was visited. In PLACE II, venues were divided into groups based on priority, and venues were sampled from the high priority groups via a predetermined structure. In the PLACE Report, FSW estimates for district where no data was collected were obtained by combining expert knowledge with prevalence estimates from other districts. For example, the FSW prevalence in Thyolo was assumed to be equal to the average prevalence of all district in the Southern Region. The complete methodology for both PLACE reports is recorded in the PLACE Report Malawi \citep{PLACEreport}.

\subsection{Auxiliary Variables}
\label{sec:aux}

Using the auxiliary variables enables us to map a more complete picture of Malawi key populations. In the PLACE report, venue weights were used to extrapolate the FSW size to the district level \citep{PLACEreport}. The proportion of venues visited and operational were used to calculate the venue weights. This venue weighting assumes that unvisited venues behave similarly to visited venues, ignoring the location and local characteristics of unvisited venues. We can improve FSW size estimates by employing a regression-based approach with auxiliary data from multiple sources which accounts for the local characteristics of all venues. The auxiliary variables that we use are the demographic and health survey (DHS) data, night-time lights (nightlight), and WorldPop.

\subsubsection{DHS Data}

\begin{figure}[!tb]
\centering
\begin{subfigure}{.245\textwidth}
  \centering
  \includegraphics[width=1.6in]{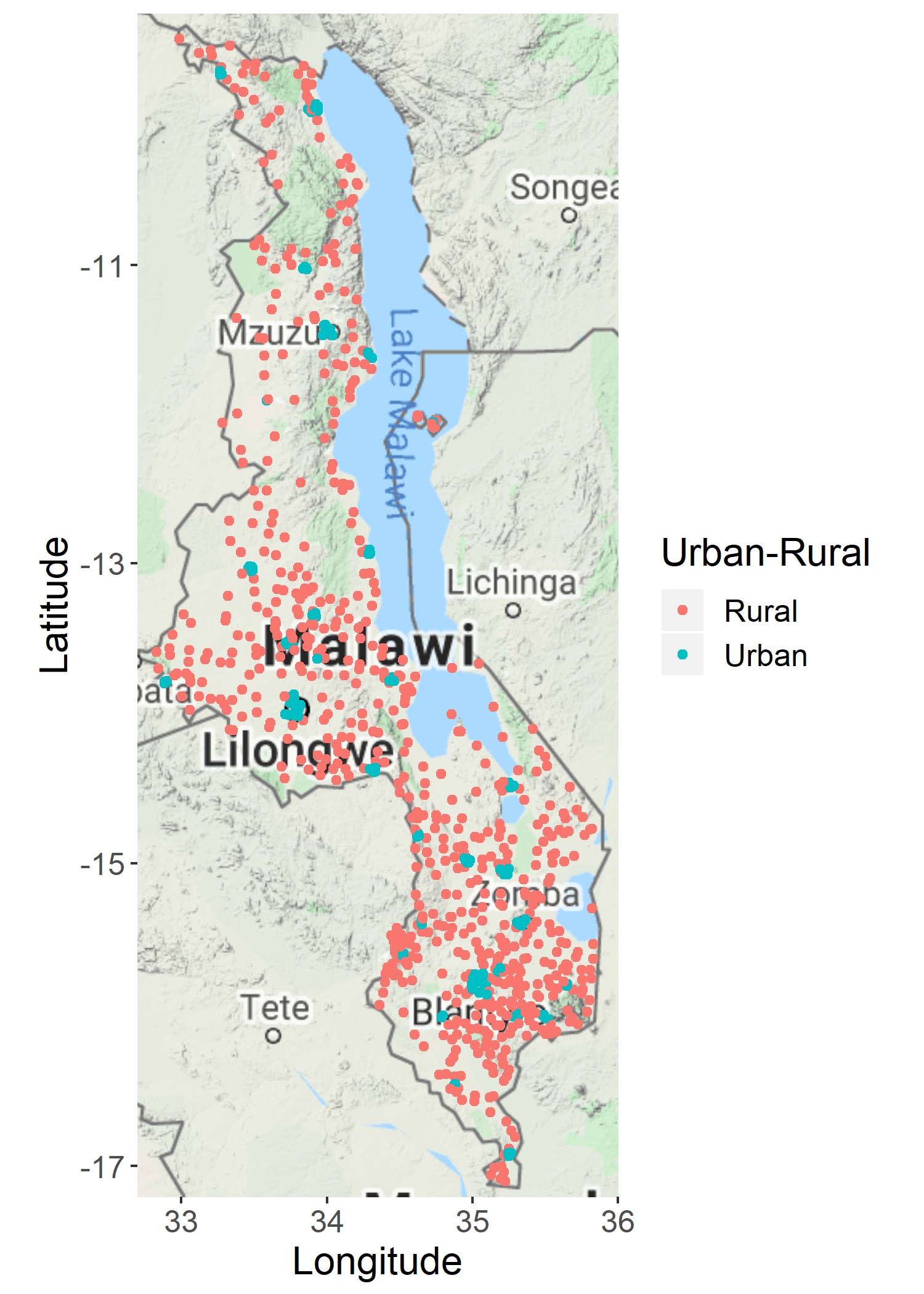}
  \caption{}
  \label{fig:mv012_val}
\end{subfigure}%
\begin{subfigure}{.245\textwidth}
  \centering
  \includegraphics[width=1.6in]{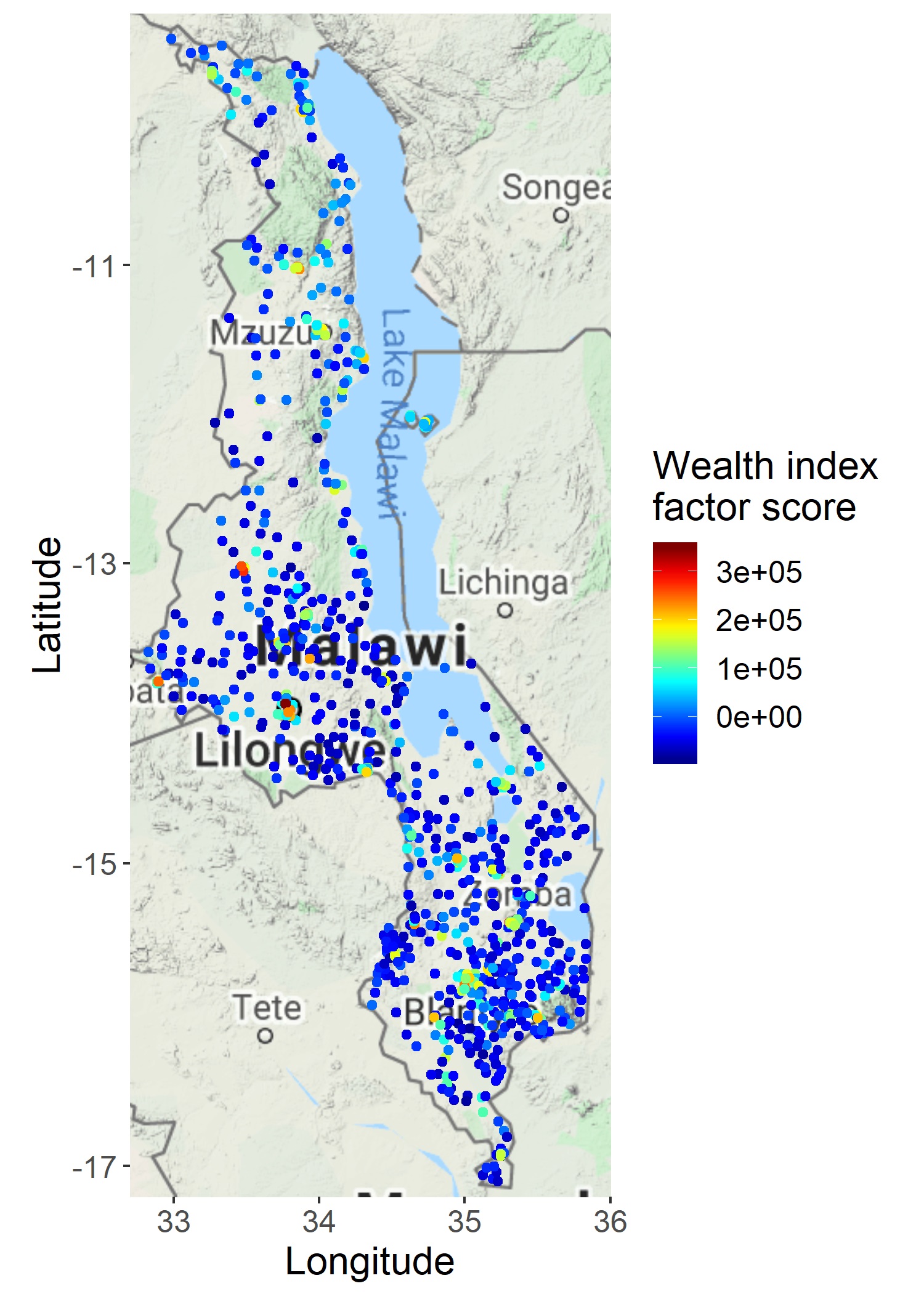}
  \caption{}
  \label{fig:mv012_var}
\end{subfigure}
\begin{subfigure}{.245\textwidth}
  \centering
  \includegraphics[width=1.6in]{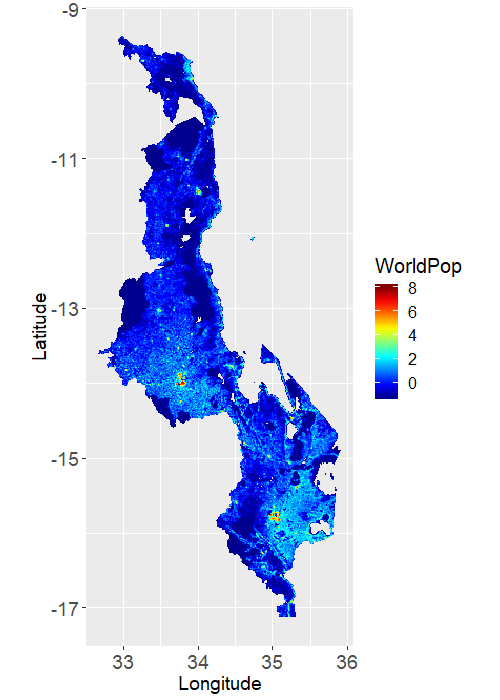}
  \caption{}
  \label{fig:worldpop}
\end{subfigure}
\begin{subfigure}{.245\textwidth}
  \centering
  \includegraphics[width=1.6in]{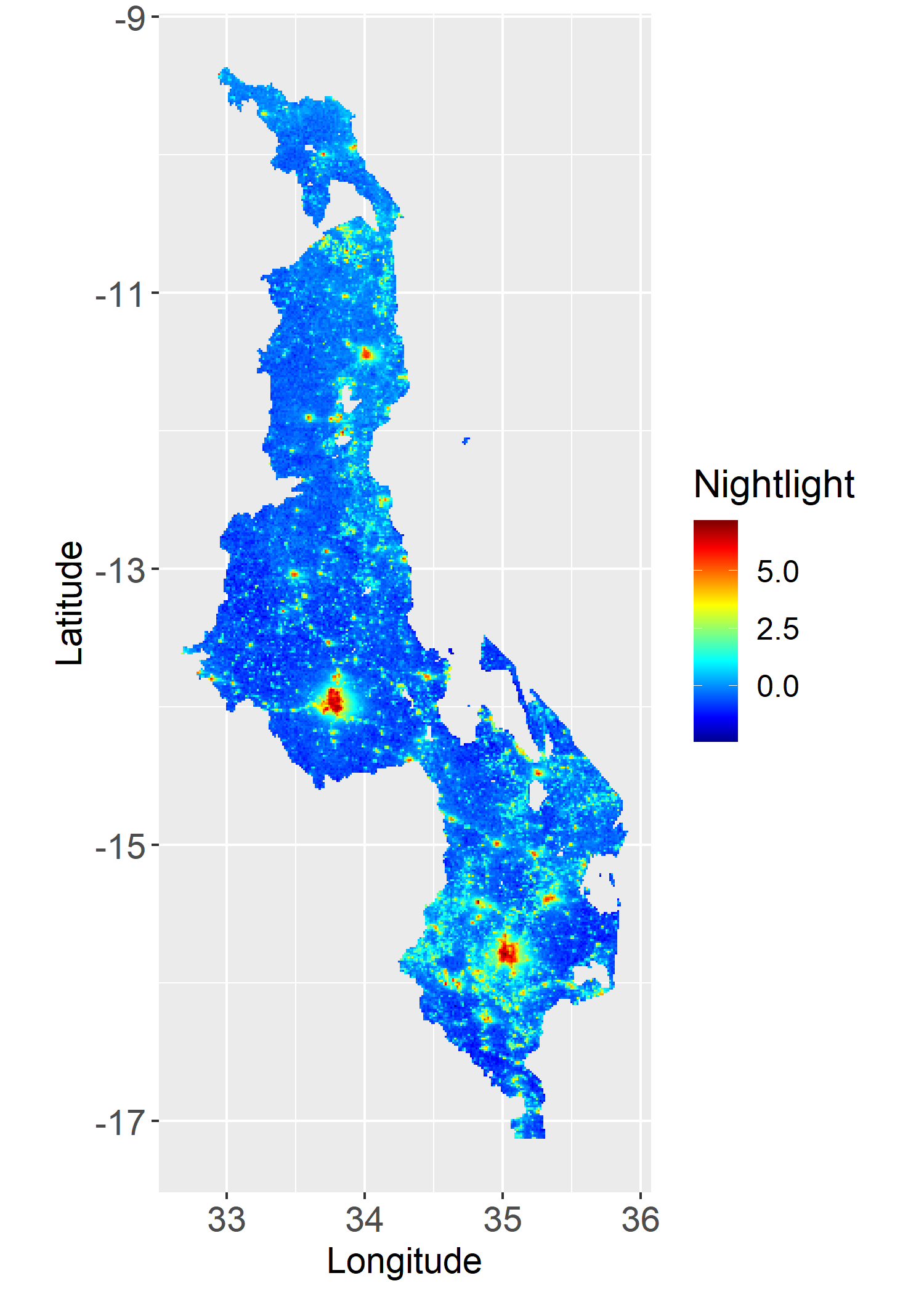}
  \caption{}
  \label{fig:nightlight}
\end{subfigure}
\caption{Illustrative maps of the auxiliary covariates. (a) shows which clusters are urban and rural. (b) shows the distribution of the mv191 (combined wealth index factor score). (c) and (d) show the scaled WorldPop and nightlight cell estimates, respectively.}
\label{fig:DHS_clust}
\end{figure}

The DHS data are the results of a survey of 26,361 household in Malawi where respondents were asked questions like annual income, age at first sex, and number of sexual partners. Respondents were surveyed between October 2015 and February 2016. DHS divided households into 850 ``clusters,'' which represent groups of household that are close in proximity. DHS perturbed the cluster locations to ensure the privacy of respondents. We treat the recorded cluster locations as the truth since the effect of the perturbations will have a much smaller effect on the estimates than the variability of the venue counts and the FSW sizes. As the data contains thousands of questions, after consulting with relevant experts, we created an initial list of covariates that we believed would be useful in modeling either the venue count or venue size. DHS clusters (as well as WorldPop and nightlight data discussed later) are shown in Figure \ref{fig:DHS_clust}.

\subsubsection{Nightlight}

The most popular FSW venues likely produce light during the night, so a measure of nightlight intensity should be an important predictor of both venue count and FSW size. We use nightlight data recorded by the Earth Observations Group at the National Oceanic and Atmospheric Administration and the National Centers for Environmental Information (NOAA/NCEI) \citep{nightlight}. The nightlight data measures the low levels of visible-near infrared radiance via satellites. The data was recorded at a roughly 450 meter resolution, resulting in $567,284$ locations of recordings from polar orbiting satellites twice a day. We use 2016 data, aggregated by month. Month averages are then averaged, yielding annual average nighttime values across Malawi. The cell-level nightlight values are extremely right-skewed, so the values were taken on the log-scale after alignment. The cell-level estimates after centering and scaling are shown in Figure \ref{fig:DHS_clust}.

\subsubsection{WorldPop}

Areas with higher populations should intuitively have more venues, and potentially more FSW per venue. In order to capture this relationship, we use the 2015 adjusted WorldPop data set, which adjusts the population estimates to match UN national estimates \citep{worldpop}. WorldPop estimates the number of persons per grid square at approximately 100 by 100 meter cells. Cells with zero WorldPop values were removed and the remaining values were taken on the log-scale. These cells typically represent inhabitable zones like forests and lakes, so we limit inference to cells that are habitable. The cell-level estimates after centering and scaling are also shown in Figure \ref{fig:DHS_clust}.

\subsection{Combining the Data}
\label{sec:comb_data}

The venues in Malawi represent a presence-only marked point process since the observations contain only the longitude, latitude, and venue size estimate of \textit{visited} venues in Malawi. However, the responses and predictors exist on different spatial levels. This issue is known as spatial misalignment, and has been discussed in detail (see \cite{mugglin2000fully}, \cite{gelfand2001change}, \cite{young2009linking}, \cite{carlin1999spatio}, \cite{banerjee2014hierarchical}.) Since we have a two-part model, we also have two spatial resolutions: cell-level data for the venue count model and venue-level data for the venue size model.

For the venue count model, we first divide Malawi into cells of approximately 1.5x1.5-km. We discuss the choice of cell size in Section 4. The areas of the cells vary between 2.318 km$^2$ and 2.392 km$^2$, which has a negligible effect on the estimates. The small differences in area arise because the distance between each degree of latitude depends on the distance to the equator. For the venue count response, we count the number of observed venues in each cell. For the DHS predictors, we perform universal kriging to estimate the value in each cell. We impose a Mat\'{e}rn covariance function on the Gaussian Process in the kriging model, and parameters are estimated using restricted maximum likelihood via the \pkg{fields} package \citep{fields}.
%The variation associated with the estimate is included in our model as measurement error.
The WorldPop and nightlight data are recorded at a much finer scale than our Malawi cells, so cell-level values are calculated by averaging the WorldPop and nightlight points which lie inside the cell. After removing the cells with WorldPop equal to zero, there are 39,312 cells for all of Malawi and 20,751 cells in PLACE II districts, of which 385 contained observed PLACE II venues.
%These values are considered to have no measurement error, unlike the DHS covariates.

%Cells with WorldPop equal to zero were removed from analysis as they represent areas where there is no population. These areas are typically lakes, mountain ranges, and forests. We remove these cells to limit our inference to cells where FSW presence is feasible.

For the venue size model, our spatial resolution is the observed venue locations. Thus, only the predictors need to be aligned since the venue size response is already known at observed venues. The procedure for the DHS predictors is the same as for the cell-level, where estimates are now obtained at venue locations via kriging. The WorldPop and nightlight values at each venue are obtained by finding the closest WorldPop and nightlight recording.

\section{Methodology}
\label{sec:NewMethod}

In this section, we introduce the approach for modeling the venue size (number of FSW at the venue-level) and venue count (number of venue at the cell-level). The venue locations comprise a presence-only data set. Our model assumes that the true cell-level venue counts follow a zero-inflated negative binomial (ZINB) distribution, and we observe a thinned version of these counts. The details of the thinning are discussed in Section \ref{sec:method_ven_count}. While analyzing spatial point patterns by dividing the domain into cells and modeling the number of occurrences in the cells is not new, we are unaware of its application to presence-only data. In this section, we first introduce the venue size and venue count models. Following the modelling choices, we discuss how the models are used to predict the number of FSW in each cell which includes a calibration step.

\subsection{Models}
\label{sec:Datas}

Here we summarize the modeling choices and further specify our choices of priors in the Bayesian framework. Both the venue counts and the venue sizes seem to follow zero-inflated distributions. We assume that given the covariates for a cell, the venue count and the venue size are conditionally independent and we assess whether or not this assumption is reasonable in subsection \ref{sec:validation}. We use the zero-inflated model described in \cite{lambert1992zero} to model both the venue counts and the FSW size. In this formulation, zeros can come from either excess zeros or from the conditional distribution, $f_c(y)$. The zero-inflated distribution is given by
\begin{equation}
    \begin{aligned}
        P(Y = 0) &= p + (1 - p)f_c(0)\\
        P(Y = y) &= (1-p)f_c(y)
    \end{aligned}
    \qquad
    \begin{aligned}
        &\text{for } y = 0 \\
        &\text{for } y > 0,
    \end{aligned}
    \label{eq:fsw_brm}
\end{equation}
where $p$ is the probability of excess zeros and the choices of $f_c(y)$ are motivated by predictive performance and are specified in Sections \ref{sec:method_ven_size} and \ref{sec:method_ven_count}.

\subsection{Venue Size}
\label{sec:method_ven_size}
We can directly model the average number of FSW at the venue-level (venue size) using both PLACE I and PLACE II data. We choose a hurdle log-normal model for the venue size model, i.e. $f_c(y)$ in Equation (\ref{eq:fsw_brm}) is given by $\log(Y_\text{size}) \sim N(\mu, \sigma_\text{size}^2)$, where $Y_\text{size}$ represents the observed venue sizes with corresponding covariate matrix $\mathbf{X}_\text{size}$. We use a logit link for $p$. Given the covariates, we assume the linear regression models:
\begin{align}
    \text{logit} (p_{i, \text{size}}) &= \alpha_{0,\text{size}} + \mathbf{X}_{i,\text{size},p}\bm{\alpha}_{1,\text{size}}  + \gamma_{d[i], \text{size},p} \\ 
    \log(\mu_{i, \text{size}}) &= \beta_{0,\text{size}} + \mathbf{X}_{i,\text{size},\mu}\bm{\beta}_{1, \text{size}}  + \gamma_{d[i], \text{size},\mu}
\end{align}
where $i$ denotes the cell number, $d[i]$ denotes the district for cell $i$, $\mathbf{X}_{i,\text{size},p}$ and $\mathbf{X}_{i,\text{size},\mu}$ are subsets of $\mathbf{X}_\text{size}$, and $\gamma$ are the random district effects. We impose weak priors on the regression coefficients, where $\alpha_0 \propto \text{logistic}(0,1)$, $\bm{\alpha}_1 \propto \mathbf{1}$, $\beta_0 \propto t_3(-2, 10)$, $\bm{\beta}_1 \propto \mathbf{1}$, $\sigma^2_\text{size} \propto t_3(0, 10)$.  For the district random effects, we impose normal priors, i.e. $\gamma_{d[i], \text{size}, \cdot} \sim N(0, \sigma^2_{\text{district},\text{size}, \cdot})$, where the $\cdot$ argument is either for $p$ or $\mu$. Additionally, $\sigma^2_{\text{district},\text{size}, \cdot} \propto \mathbf{1}$.

We choose to model venue size via a hurdle log-normal distribution based on both visual diagnostics, posterior predictive p-values, and leave-one-out (loo) cross-validation. The hurdle log-normal distribution offers much better results for these metrics than a ZINB distribution, despite the venue sizes being counts. Because of the large number of zeros, non-zero-inflated models do not perform well.

\subsection{Venue Count}
\label{sec:method_ven_count}
We first assume that the true cell-level venue counts, $\tilde{\bm Y}$, follow a ZINB distribution and the observed venues $\bm{Y}$ follow a $\pi$-thinning of $\tilde{\bm Y}$, i.e.
\begin{equation}
\begin{split}
        Y_i &= \sum_{k=1}^{\tilde{Y}_i} I_{i,k},\\   
        %f_c(\tilde{y}) &= \binom{\tilde{y} + \phi - 1}{\tilde{y}}\left(\frac{\mu}{\mu  + \phi} \right)^{\tilde{y}} \left(\frac{\phi}{\mu + \phi} \right)^\phi, \\
        \tilde{Y}_i &\sim ZINB(p_i, \mu_i, \phi) \\
        I_{i,k} &\sim Bern(\pi_{d[i]})
\end{split}
\label{eq:nb}
\end{equation}
where $I_{i,k}$ indicates whether venue $k$ in cell $i$ was sampled and $\pi_{d[i]}$ is the sampling probability for the district for cell $i$. Here, we use $ZINB(p_i, \mu_i, \phi)$ to denote the zero-inflated negative binomial distribution with probability of excess zero $p_i$, conditional mean $\mu_i$, and conditional over-dispersion $\phi$, where the negative binomial parameterization we consider is
\begin{equation}
    f_c(\tilde{y}_i) = \binom{\tilde{y}_i + \phi - 1}{\tilde{y}_i}\left(\frac{\mu_i}{\mu_i  + \phi} \right)^{\tilde{y}_i} \left(\frac{\phi}{\mu_i + \phi} \right)^\phi.
\end{equation}
This $\pi$-thinning, also known as binomial subsampling, assumes that each venue is sampled independently according to the sampling effort in each district \citep{puig2007characterization}.

In order to show that the distribution of $\bm{Y}$ also follows a ZINB distribution where only the mean of the negative binomial component is multiplied by $\pi_{d[i]}$, we present the following two propositions.

\begin{lemma}
    Let a random variable $X \sim NegBin(\text{mean} = \mu, \text{dispersion} = \phi)$, as defined in (\ref{eq:nb}). Let $Y = \sum_{k=1}^{X} I_k$, where $I_k \sim Bern(\pi)$, the $\pi$-thinning of $X$. Then, $Y \sim NegBin(\text{mean} = \pi \mu, \text{dispersion} = \phi)$.
\end{lemma}

\begin{lemma}
    Let a random variable $\tilde{Y} = (1-Z)X$ be any zero-inflated discrete random variable, where $Z \sim Bern(p)$ and $f_c(x)$ represents the p.m.f. of $X$. Let $X^* = \sum_{i=1}^X I_{x,i}$ denote the $\pi$-thinning of the count distribution, where $I_{x,i} \sim Bern(\pi)$. If $Y = \sum_{i = 1}^{\tilde{Y}} I_{y,i}$, where $I_{y,i} \sim Bern(\pi)$, then $Y \overset{D}{=} (1 - Z)X^*$.
\end{lemma}

Another way of stating Proposition 2 is that the thinning operator ignores the zero-inflated component. With these two propositions, we arrive at the main proposition for our model.

\begin{lemma}
    Let $\tilde{Y}$ be a zero-inflated negative binomial random variable, with negative binomial mean $\mu$ and dispersion $\phi$, i.e. $\tilde{Y} \sim ZINB(p, \mu, \phi)$. If $Y = \sum_{k=1}^{\tilde{Y}} I_k$, where $I_k \sim Bern(\pi)$, then $Y \sim ZINB(p, \pi \mu, \phi)$.
\end{lemma}

All proofs are shown in the Appendix. Using Proposition 3, $Y_i \sim ZINB(p_i, \pi_{d[i]} \mu_{i, \text{count}}, \phi)$, where $Y_i$ is our observed venue count for cell $i$. In standard generalized linear model terms, this is equivalent to adding an offset of $\log(\pi_{d[i]})$ to the linear predictor, providing an easy way to fit the model using common statistical software. Because the sampling probability in each district, $\pi_{d[i]}$, are known, we can estimate all parameters in the original model directly by fitting an offset model. Thus, our model for the observed venue counts, probability of excess zeros $\mathbf{p}_\text{count}$, and mean of the conditional distribution $\bm{\mu}_\text{count}$ are given by
\begin{equation}
\begin{split}
    Y_i &\sim ZINB(p_{i, \text{count}}, \mu_{i, \text{count}}, \phi) \\
    \text{logit} (p_{i, \text{count}}) &= \alpha_{0,\text{count}} + \mathbf{X}_{i,\text{count},p} \bm{\alpha}_{1,\text{count}} + \gamma_{d[i], \text{count},p} + \eta_{i, \text{count}, p} \\ 
    \log(\mu_{i, \text{count}}) &= \beta_{0,\text{count}} +  \mathbf{X}_{i,\text{count},\mu} \bm{\beta}_{1,\text{count}} + \gamma_{d[i], \text{count},\mu} + \log(\pi_{d[i]}),
\label{eq:nb2}
\end{split}
\end{equation}
where ${\bm \eta} \sim MVN(0, k_p(x))$. Based on the data and loo cross-validation, a spatial error term ${\bm \eta}$ is placed on the Bernoulli model but not on the negative binomial model. We impose the same priors on the regression parameters and the random effects. For the shape of the negative binomial distribution, we impose a gamma prior, i.e. $\phi \sim gamma(0.01, 0.01)$. For the spatial term ${\bm \eta}$, we choose the exponentiated-quadratic kernel $k(x_i, x_j) = \sigma^2_{gp} \exp(-||{x_i - x_j}||^2/ (2 l_\text{scale}^2))$ and impose a flat prior on $\sigma^2_{gp}$ and an informative inverse-gamma prior on $l_\text{scale}$ as $l_\text{scale} \propto inv-gamma(0.976289, 0.008892)$, where the hyperparameters were estimated from the data.

One key advantage of our model is that traditional GLM regression diagnostics can be used, a significant limitation of other presence-only models. Furthermore, our cell-level approach can be extended to other distributions, like the zero-inflated Poisson, or non-zero inflated Poisson and negative binomial distributions, for example. The method requires only knowing the distribution of the thinned process and being able to estimate the parameters of the original process using only the thinned process.

\subsection{Prediction and Calibration}
To predict the number of FSW in each cell, we fit a venue size model and a venue count model independently. The venue size model estimates the number of FSW at a venue via a hurdle log-normal distribution. This model is fit using the observed venue sizes. Using cell-level covariates, we predict the venue size for venues in all cells. The calibrated venue count model estimates the number of venues in a cell in three steps: (1) fit a ZINB model using observed venue counts, accounting for the thinning process in each district via an offset, (2) predict the venue count for all cells using the fitted model, and (3) calibrate the predicted venue counts in each district to match the known district-level venue counts. After modeling venue size and count, we predict the cell-level FSW sizes by multiplying the calibrated venue counts by the predicted venue sizes.

Predictions of the true venue counts can be made by setting $\pi_{d[i]} = 1$ for all cells. For districts where district-level venue counts are known, we calibrate the predictions from each posterior sample of the venue count model so that the predicted venue count equals the known venue count at the district-level. For each district, this is done in two steps for each set of posterior samples: (1) obtain district-level venue estimates for each posterior sample by summing the predicted values for each cell with $\pi_{d[i]} = 1$, and (2) multiply each predicted value for the posterior sample by the calculated scaling factor $\lambda$, so that the sum of predicted values equals the known count. However, we perform no calibration for the remaining districts, as no information is available. Note that we simulate district random effects when they were not estimated from the model. These random effects are the PLACE I districts and remaining unvisited districts in the venue count model and the unvisited districts in the venue count and venue size models.

Note that step (1) already assumes the expected number of venues in each district is equal to the district-level venue counts provided by the PLACE report for districts included in the venue count model (only PLACE II districts). Step (3) then calibrates the venue counts for the remaining districts where district-level venue counts are known. To remain consistent across districts, we calibrate the venue counts for all districts where the counts are known. If the district level venue counts are believed to have measurement errors, then the exact matching is not needed and step (3) could be omitted entirely or relaxed. For Malawi, the PLACE team interviewed thousands of community informants to find all possible venues, so we believe the district-level venue counts very accurate, and thus adopt the exact matching approach.

\section{Model Validation}
\label{sec:validation}

In this section, we check our model assumptions for the PLACE data, and where that is impossible, we use a simulation study to explore the behavior of the models when those assumptions are violated. We have four main assumptions which we check in the following order: (1) the true venue counts follow a zero-inflated negative binomial distribution, (2) venue size and venue count are conditionally independent, (3) the cell resolution provides accurate results, and (4) venues were sampled uniformly within each district. We can check the first two assumptions directly for our PLACE data, but we rely on a simulation study to evaluate the effect of cell-resolution and non-uniform sampling.

\subsection{Real Data Validation}
Checking the performance of presence-only models using real data is difficult because the truth is typically unknown. However, the PLACE team visited almost every venue in three districts, Mchinji, Mwanza, and Neno, which provides us with a validation data set where almost all true venue counts are observed. The number of observed venues against the number of operational venues for the complete districts are 98/102, 75/78, and 64/65, respectively. We refer to these three districts as the complete districts, even though a few venues were not visited. Note that even though almost all venues were visited, the venue sizes are not available at all venues, so the district-level FSW sizes are still unknown. We use these complete districts to first show the appropriateness of our assumed ZINB distribution for the venue counts, and then to show the conditional independence between the venue counts and the venue sizes.

\subsubsection{Distribution of True Venue Counts}
We first check that the true venue counts follow a ZINB distribution by fitting a non-thinned ZINB model with the venue counts from the complete district. We rely on posterior predictive p-values to verify the goodness of fit, and these results can be found in Table \ref{tab:complete_ppp} in the Appendix. More details about posterior predictive p-values and test statistics can be find in Section \ref{sec:RealData} where we introduce the p-values for all of Malawi. Based on these values, we find the ZINB distribution provides a very reasonable goodness of fit for the complete districts. Visual diagnostics similar to those in Figure \ref{fig:diags} were also considered for the complete districts, but are not included here as there is no clear evidence that the models do not adequately fit the data.

\subsubsection{Model Independence}
\begin{figure}[!tb]
    \centerline{\includegraphics[width=0.6\textwidth]{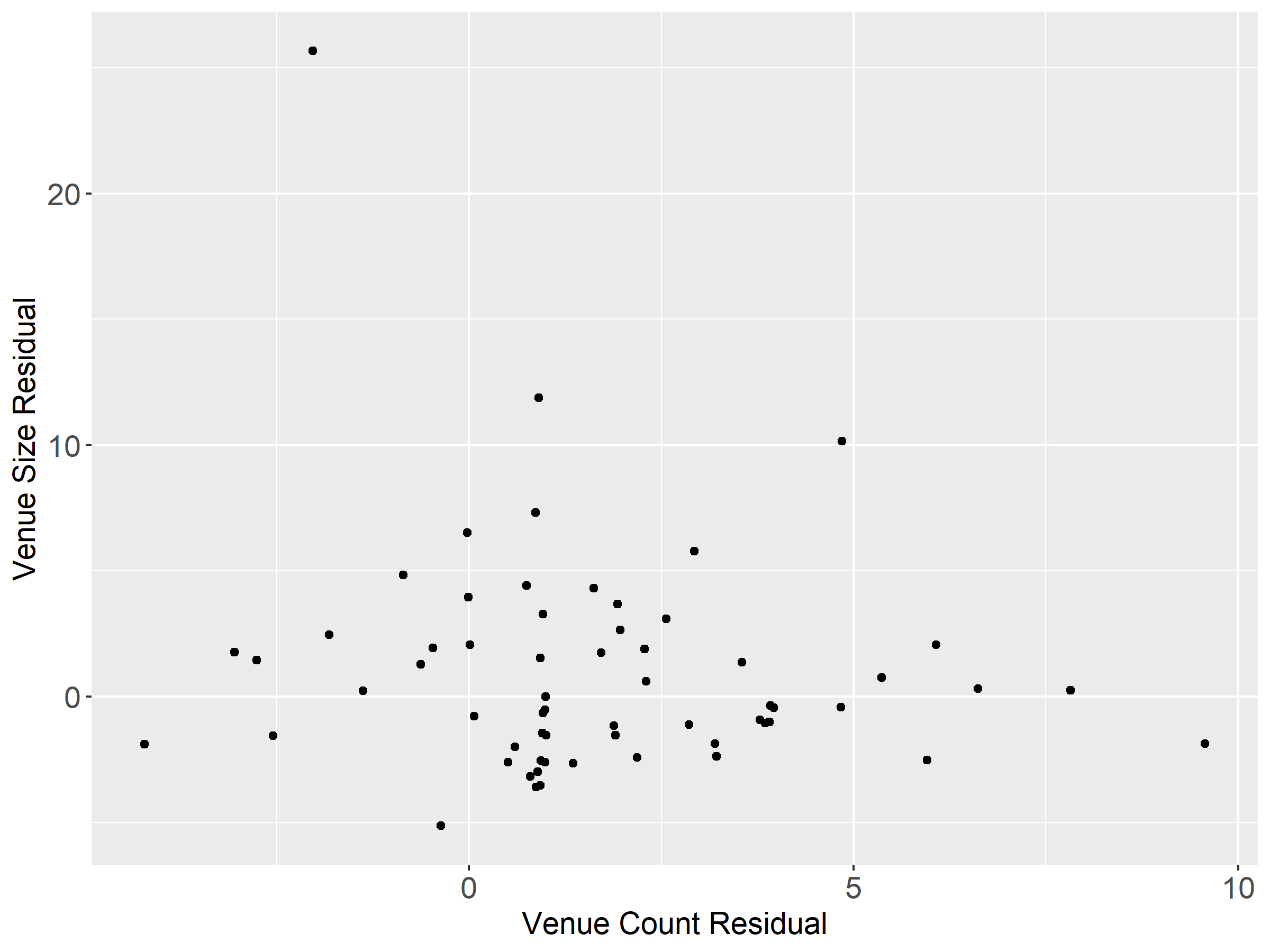}}
    \caption{A scatter plot of venue size residuals and venue count residuals.}
    \label{fig:venue_ci}
\end{figure}

Next, we provide evidence for the independence between the venue counts and venue sizes by examining the cell-level residuals. For the venue count model, the residuals are defined as the observed number of venues in cell $i$ minus the predicted number of venues. For the venue size, the residuals are the average venue size in cell $i$ minus the predicted venue size. However, since true-data is generally unknown, we limit our analysis to only the three complete districts and only the cells with at least one venue. Cells with zero venues are excluded because there is no average venue size. The scatter plot of the venue size residuals and venue count residuals is shown in Figure \ref{fig:venue_ci}. The residuals do not appear to have any strong correlation structure, suggesting the conditional independence assumption is reasonable.

\subsection{Simulation Study}
To address the final two assumptions, we rely on a simulation study. We first show that cell resolution does affect total FSW size estimates, but it is difficult to determine the optimal cell resolution using residual diagnostics. We then show that if venues are sampled proportionally to a covariate, total FSW size is again biased, but including the relevant covariate helps mitigate most of the bias. We now introduce our simulation design. For our simulation study, we control three factors: cell resolution, (non)uniform sampling, and whether all covariates related to nonuniform sampling are included in the regression model. 

\emph{Design of cell resolutions:} Our spatial domain for the simulations is the unit square. Specifically, we first simulate a spatial grid of covariates, divide the domain into a grid of 45 by 45 equal sized cells, yielding 2025 cells. In each cell, we simulate the counts, locations, and sizes of venues according to a ZINB distribution for the counts and a hurdle log-normal distribution for the sizes. Simulation parameters are chosen to resemble the PLACE data. These 2025 cells represent the ``true resolution'' since the venue counts at this resolution are simulated from a ZINB distribution. We then divide the domain into either 484 (large) cells, 5041 (small) cells, or use the true cells, and find the covariates in each cell. The large cells correspond to a grid of 22 by 22 equal sized cells and the small cells to 71 by 71 equal sized cells (relating these cell sizes back to the Malawi study, these are roughly equivalent to considering $3 \times 3$-km or $1 \times 1$-km cells instead of our $1.5 \times 1.5$-km cells).

\emph{Design of (non)uniform sampling:} We thin the venues either uniformly (half of all venues are sampled with equal probability) or nonuniformly (half of all venues are sampled where the probability of sampling a venue is strongly correlated with one of the covariates). The observed venue counts are found for each cell resolution, our thinned ZINB model is fit to the counts and calibrated, and a hurdle log-normal model is fit to the sizes. For the fitting, we either fit the model with all covariates (all covariates) or exclude the covariate correlated to the sampling probabilities (missing covariate). 

\begin{figure}[!tb]
\centering
\begin{subfigure}{.245\textwidth}
  \centering
  \includegraphics[width=1\linewidth]{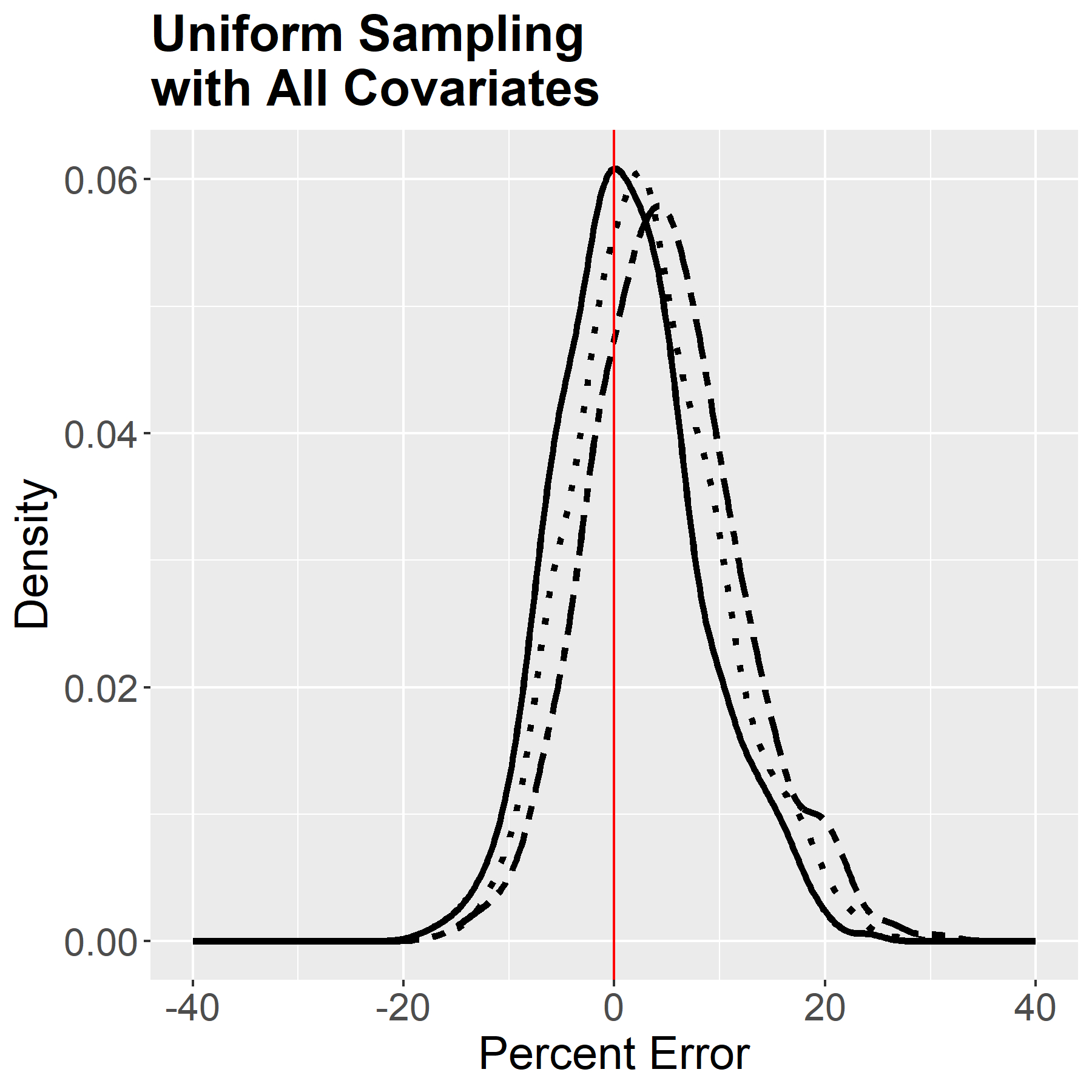}
  \caption{}
  \label{fig:cell1}
\end{subfigure}%
\begin{subfigure}{.245\textwidth}
  \centering
  \includegraphics[width=1\linewidth]{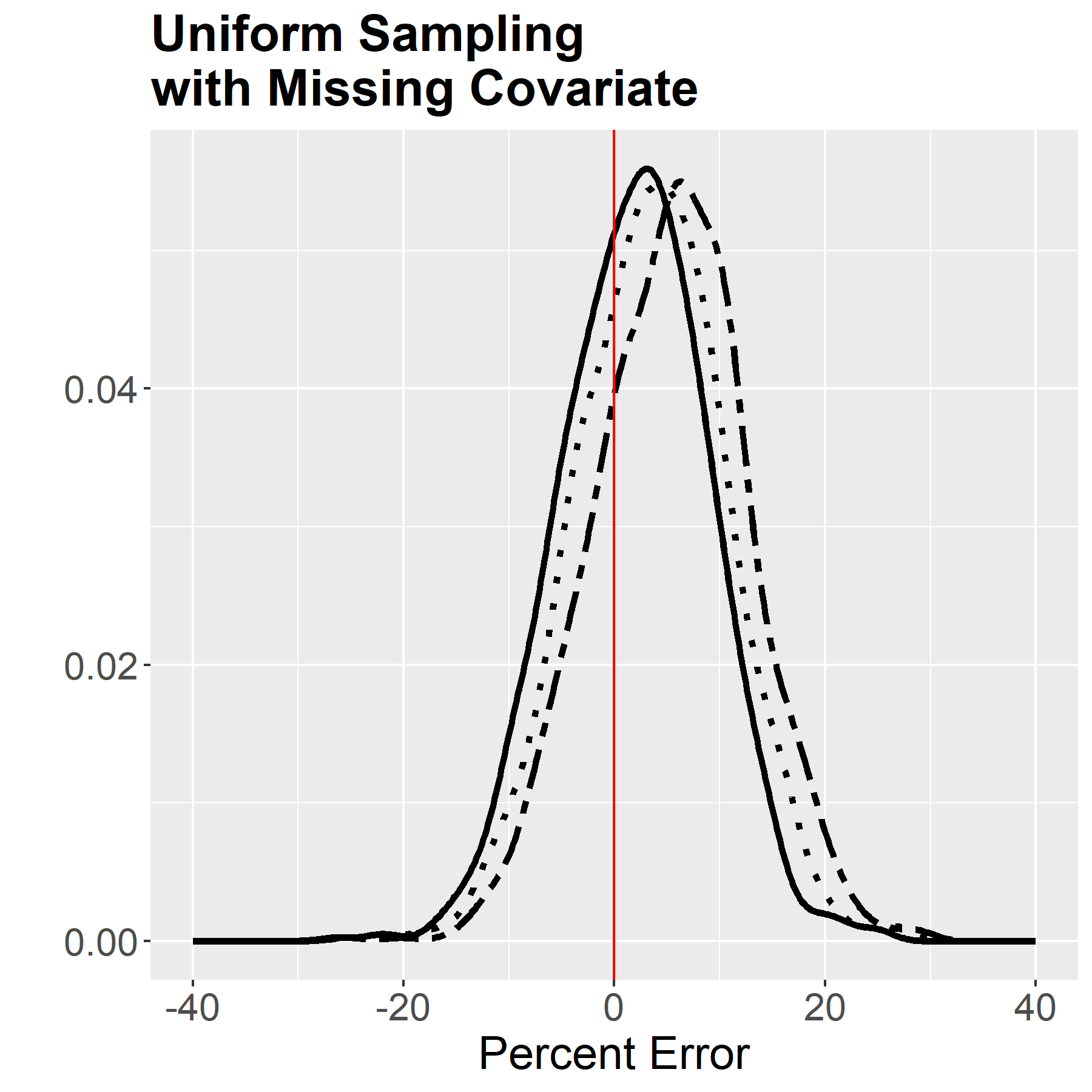}
  \caption{}
  \label{fig:cell2}
\end{subfigure}
\begin{subfigure}{.245\textwidth}
  \centering
  \includegraphics[width=1\linewidth]{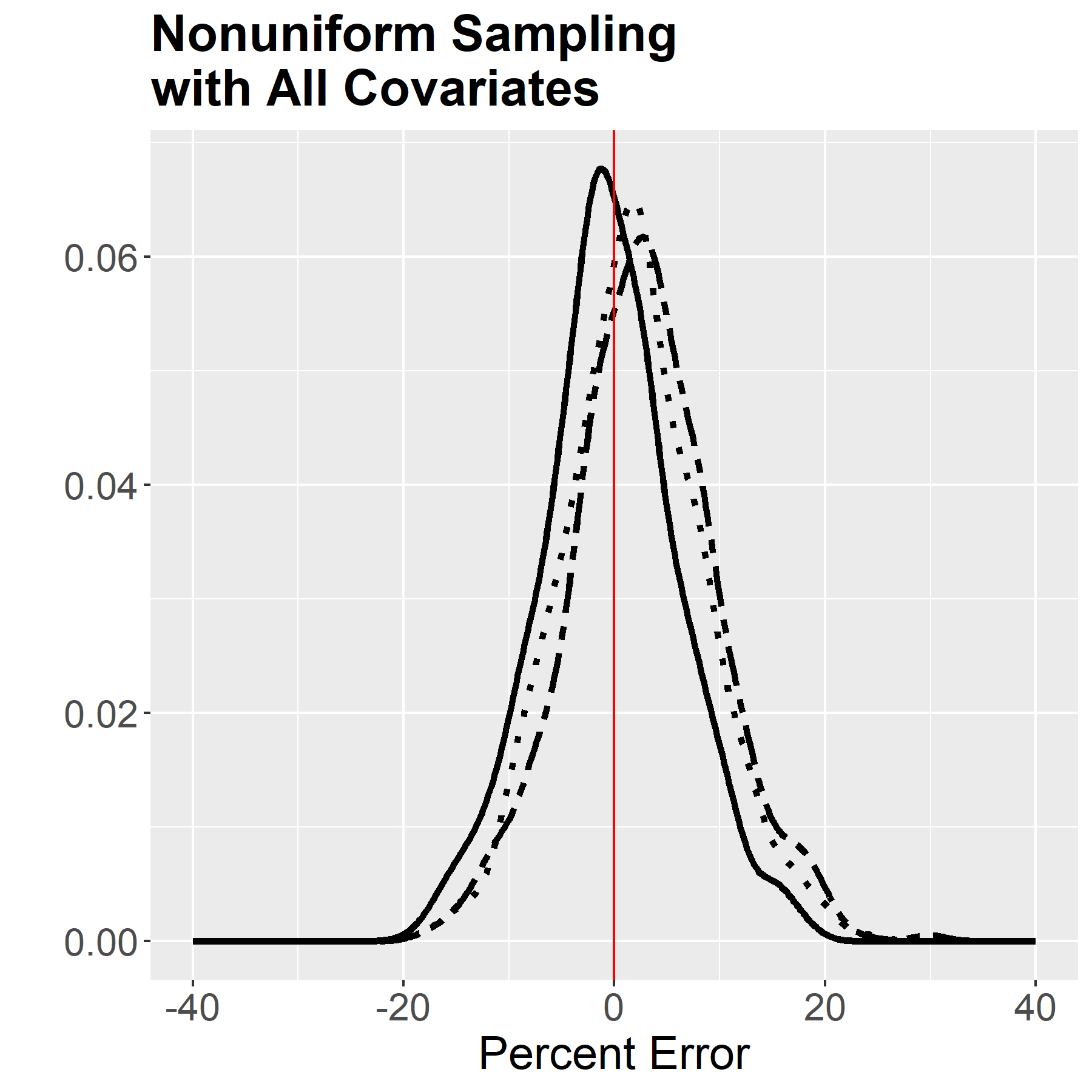}
  \caption{}
  \label{fig:cell3}
\end{subfigure}
\begin{subfigure}{.245\textwidth}
  \centering
  \includegraphics[width=1\linewidth]{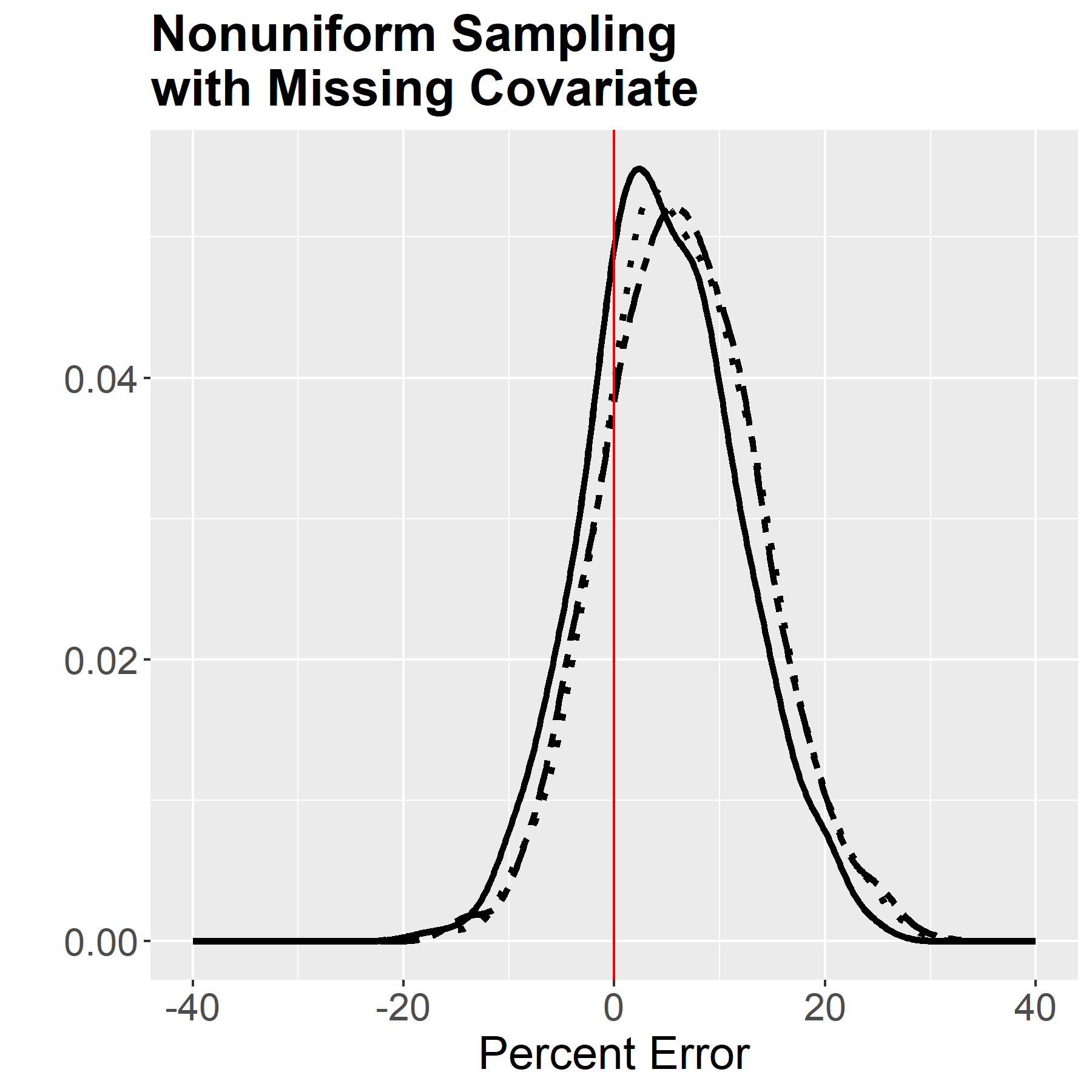}
  \caption{}
  \label{fig:cell4}
\end{subfigure}
    \caption{Distribution of total FSW estimates under different model assumptions and cell resolutions. We consider uniform sampling of venues with all covariates in the regression model (a), uniform sampling with a missing covariate (b), nonuniform sampling with all covariates (c), and nonuniform sampling with a missing covariate (d). The solid line corresponds to the true cell resolution, the dotted line to small cells, and the dot-dash line to large cells. Percent error is defined as $(Y - \hat{Y})/Y$, where $Y$ is the true total FSW size.}
    \label{fig:sim_perf}
\end{figure}

Total FSW size is found by predicting the count and size to each cell for each posterior sample, calibrating the counts, summing the cell-level FSW estimates across all cells, and finding the mean of each predicted total FSW size. This process is repeated 1000 times. The results for the simulations are shown in Figure \ref{fig:sim_perf}, where the different cell sizes are shown as different line types, the different model assumptions are provided in the titles, and the percent error is defined as $(Y - \hat{Y})/Y$, where $Y$ is the true total FSW size

\subsubsection{Cell Resolution}

We first show that cell resolution affects total FSW size estimates. We can see in subplot (a) that under the ideal scenario when sampling is uniform and all covariates are included in the regression model, the percent error is centered around zero. Under all scenarios, the cell resolution does influence the results, but the distribution of posterior-means are mostly overlapping. Based on these simulations, there also does not appear to be any systematic bias based on cell-size. For scenario (a), the average percent errors for the same, small, and large cell resolutions are 1.39\%, 3.19\%, and 4.91\%, respectively. The difference between the average percent errors is similar for the other scenarios. In all cases, the residual diagnostics of all cell resolutions appear to provide good fit, with no model obviously performing better than another, suggesting that it is very difficult to choose the best cell resolution based only on the data.

\subsubsection{Uniform Sampling}
Finally, we conclude from Figure \ref{fig:sim_perf} that while nonuniform venue sampling biases total FSW estimates, including the covariate related to sampling helps mitigate this bias. Comparing subplots (a) and (c), we can see that nonuniform sampling does introduce slight bias. For subplot (d), excluding the covariate related to sampling further biases the results. While results are best when sampling is uniform and all covariates are included in the model, the errors introduced by violations of these assumptions are still relatively small. For the ``same'' cell resolution, the average percent errors for each assumption scenario are 1.39\% (a), 1.89\% (b), -0.45\% (c), and 4.56\% (d), clearly showing the larger bias introduced by excluding the sampling covariate, but the differences between the average errors are relatively small compared to the variation introduced by observing only half of the venues.

\section{Estimation of FSW in Malawi}
\label{sec:RealData}

Here we apply our model to the PLACE data and map the distribution of FSW in Malawi. We first divided Malawi into cells. The cell size should be small enough such that the behavior inside the cells are homogeneous while large enough that computation is feasible. To balance these two factors, we divided Malawi into cells of approximately 1.5x1.5-km. For the venue size model, we used all visited venues from both PLACE I and II studies. Due to the convenience sampling for venue locations in PLACE I, only PLACE II venues are used to model venue count, excluding Mzimba. Mzimba was removed because the largest city of Mzimba was visited in PLACE I, while only the rest of Mzimba was visited in PLACE II. Additionally, 15 venues with more than 100 recorded FSW were excluded from the venue size model because in all cases, the recorded number of women present at a busy-time was less than 100, clearly showing the very large counts are unreliable. This process leaves us with 20,751 cells for the venue count model and 2,541 venues for the venue size model.

All analysis was done using \textsf{R} \citep{rlang}. Covariates were standardized prior to fitting. Spatial kriging was done via the \pkg{fields} package \citep{fields}. Bayesian modeling was implemented using the Stan probabilistic programming language \citep{stan} and the \pkg{brms} package \citep{brms}. Since the number of samples was fairly large, we fit an approximate Gaussian Process with 5 basis functions and a multiplicative constant of $5/4$ for $\eta_{i,\text{count},p}$, as suggested in the \pkg{brms} package. The multiplicative constant defines the range over which predictions from the Gaussian Process should be computed, and complete details of the approximate Gaussian Process used are found in \cite{riutort2020practical}.

\begin{table}[!tb]
\centering
\caption{Posterior predictive p-values for Malawi.}
\begin{tabular}{lcc}
\hline \hline
\textbf{Test Statistics}  & \textbf{Venue Count} & \textbf{Venue Size}  \\ \hline
Positive Mean & 0.487  & 0.482 \\
Positive Standard Deviation & 0.680  & 0.523 \\
Over-dispersion index & 0.725 & 0.548 \\
Maximum & 0.855  & 0.777 \\
Proportion of Zeros &  0.521 & 0.507 \\ \hline
\end{tabular}
\label{tab:ppp-values}
\end{table}

\begingroup
\setlength\tabcolsep{0.15in}
\def\arraystretch{1}
\begin{table}[!t]
\centering
\caption{Table of predictors in the final models. Districts are included as random effects in all models. The $-$ and $+$ indicate the direction of significant coefficients for each model at the significance level 0.05, $\bullet$ indicates the predictor was included for predictions but its coefficient was not significant, and a blank space indicates the predictor was not included in the model.}
\begin{tabular}{ m{5em}  m{12em} m{2.5em}  m{2.5em}  m{2.5em} m{2.5em} }
\hline \hline
\textbf{Predictor Label}  & \textbf{Meaning} & $\bm{\mu}_\text{size}$ & $\mathbf{p}_\text{size}$ & $\bm{\mu}_\text{count}$ & $\mathbf{p}_\text{count}$  \\ \hline
v155 &  Literacy   &  &   & $\bullet$ &  $\bullet$ \\
mv201 & Total children ever born &  & - &  &  \\ 
mv167 & Times away from home in last 12 months & $\bullet$ &  &  & $\bullet$ \\
Built & Built-up index. Higher values roughly correspond to more buildings &    & + & $\bullet$ & + \\
hivpos  & Percent HIV positive  & $\bullet$  &  &  &  \\
WorldPop  & 2015 population estimate & $\bullet$ & - & + & - \\
nightlight & Night-time light activity  & + & -  &  & - \\ \hline
\end{tabular}
\label{tab:FSW_cov}
\end{table}
\endgroup

We again investigate model fit via posterior predictive p-values and visual diagnostics. These p-values are shown in Table \ref{tab:ppp-values} for both the venue count and venue size model and visual posterior checks are shown in Figure \ref{fig:diags} in the Appendix. For a given test statistic $T(y)$ for data $y$, the posterior predictive p-value is defined as $P(T(y^{\text{rep}}) > T(y) | y)$, where $y^{\text{rep}}$ represents the potential replication of the data \citep{meng1994posterior, gelman2013two}. Extremely large or small posterior predictive p-values might indicate violations of model assumptions. The test statistics we consider are the positive mean ($T(y) = E(y | y > 0)$), positive standard deviation ($T(y) = SD(y | y > 0)$), the over-dispersion index ($T(y) = Var(y) / E(y)$), the maximum ($T(y) = \text{max}(y)$), and the proportion of zeros ($T(y) = P(y = 0)$). These test statistics were chosen because they check the main aspects of our zero-inflated distributions. Based on these metrics, the observed venue counts closely followed a ZINB distribution. Similarly, the hurdle log-normal distribution outperformed other standard models we considered like the ZINB and the values indicate no clear lack of fit.

Table \ref{tab:FSW_cov} summarizes the variables selected in each model, where $\bm{\mu}_\text{size}$ represents the mean of the log-normal distribution on the log scale for the venue size, $\bm{\mu}_\text{count}$ the mean of the negative binomial distribution for the venue count, and $\mathbf{p}_\text{size}$ and $\mathbf{p}_\text{count}$ for mean of the Bernoulli distribution for the venue size and count models, respectively. To avoid overfitting, variables were selected by removing the least significant variable sequentially from saturated occurrence and abundance models. The collection of nested models were compared using posterior predictive p-values, leave-one-out predictive error, and visual diagnostics. The selected model performed the best with respect to leave-one-out predictive error and the visual diagnostics and posterior predictive p-values did not indicate any poor model fit. Note that our goal is not testing the significance of associations but providing the spatial distribution of the venues and FSWs. The coefficients represent the conditional relationship given other covariates instead of the marginal relationship.

\begin{figure}[!t]
    \centerline{\includegraphics[width=0.8\textwidth]{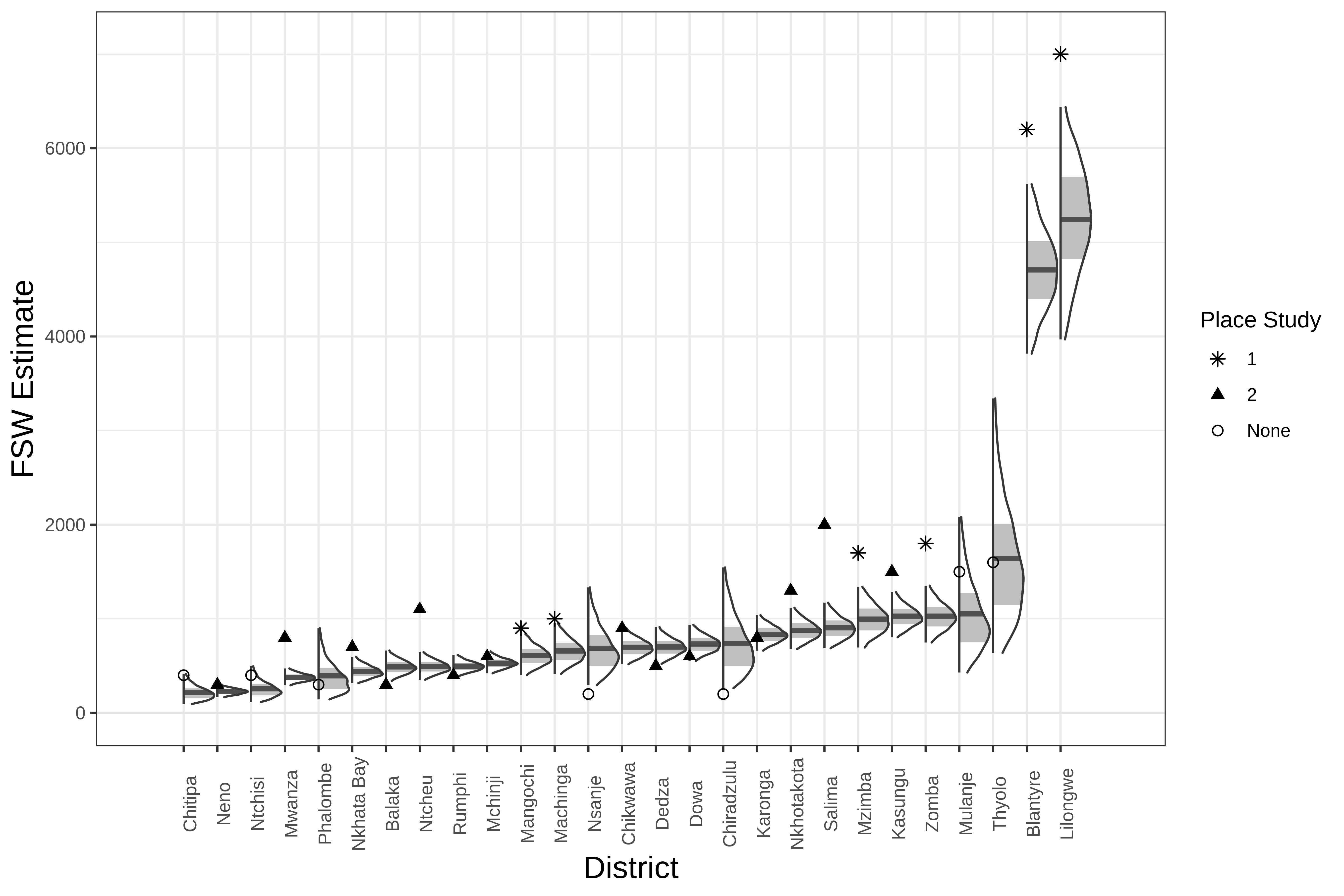}}
    \caption{District-level FSW size 95\% credible intervals for all districts in Malawi. The light gray area indicates the 50\% credible interval and the dark gray lines indicate the posterior mean. Stars, triangles, and circles indicate the PLACE Report estimates for PLACE I districts, PLACE II districts, and districts in neither PLACE study, respectively.}
    \label{fig:fsw_dist}
\end{figure}

The district-level FSW size estimates are shown in Figure \ref{fig:fsw_dist} as posterior densities. Stars, triangles, and circles indicate the PLACE Report estimates for PLACE I districts, PLACE II districts, and districts in neither PLACE study, respectively. There is no FSW estimate from the PLACE Report for Likoma. The population of Likoma is extremely small compared to the other districts, so we do not include credible intervals for Likoma in Figure \ref{fig:fsw_dist}. Not all 95\% credible intervals cover the PLACE report district estimates, although most estimates are very close. Note, however, that the PLACE estimates are not gold standards for the district-level FSW estimates. We expect the PLACE district estimates to be larger than our estimates because the PLACE method likely oversamples from large venues in populated areas. We find this to be the case for many, but not all districts. The random effects were not estimated for PLACE I districts in the venue count model or for districts in neither PLACE study in the venue size and count models, so these districts have larger credible intervals due to simulating the random effects.

\begin{figure}[!t]
    \centerline{\includegraphics[width=0.8\textwidth]{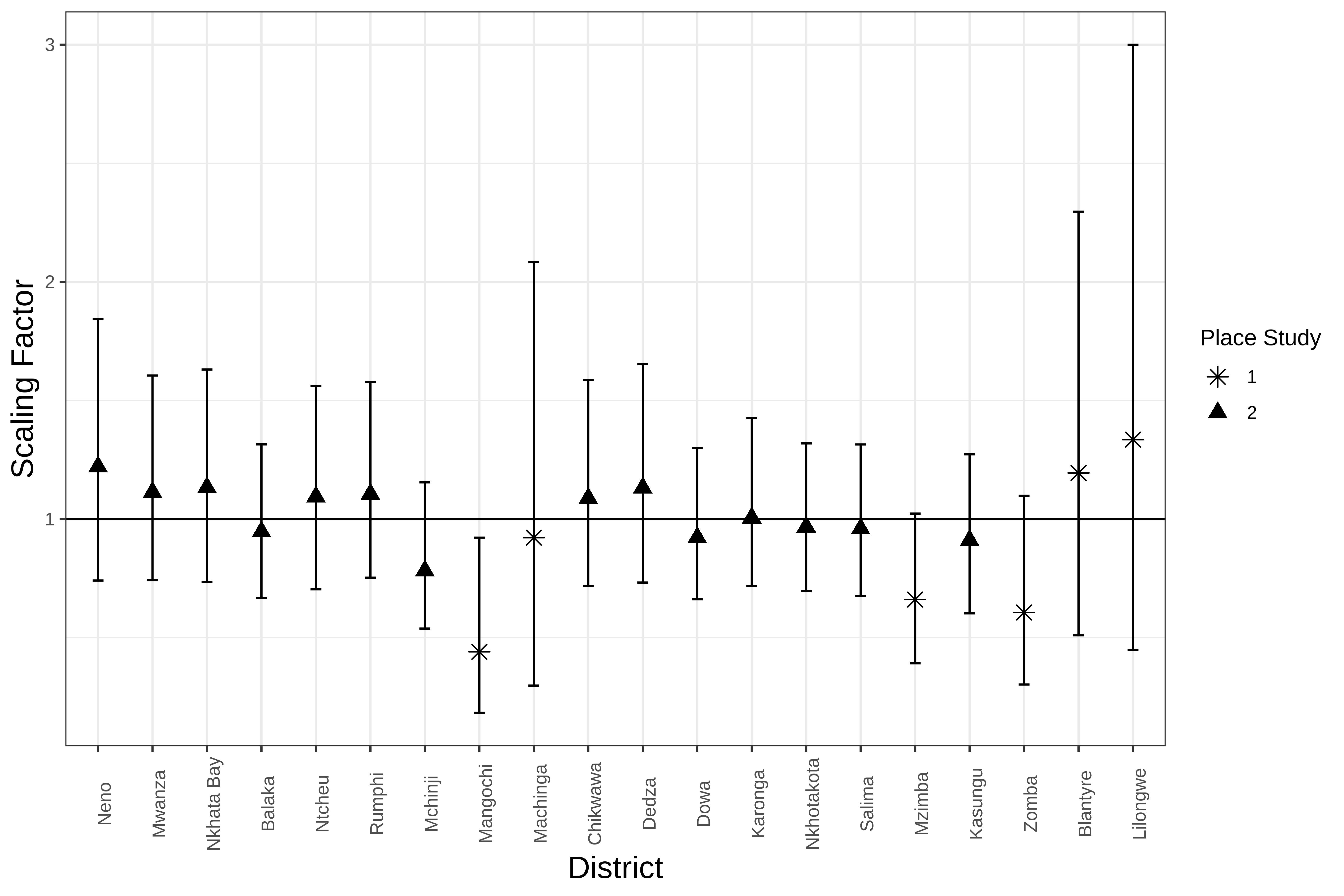}}
    \caption{Scaling factors $\lambda$ for each districts. Stars, and triangles indicate the mean $\lambda$ for PLACE I and PLACE II districts, respectively (no scaling factors can be estimated for other districts). The error bars indicate the lower and upper 95\% quantiles of the estimated $\lambda$. A flat line at $\lambda = 1$ is shown for reference.}
    \label{fig:scaling_dist}
\end{figure}

We also present the estimated scaling factors $\lambda$ for each district in Figure \ref{fig:scaling_dist}. The $\lambda$ represents the discrepancy between the known total number of venues and the estimated total number of venues in each district, where $\lambda = 1$ corresponds to perfectly estimated venue counts and $\lambda > 1$ corresponds to underestimated venue counts. For districts where venue count data was included in the model (PLACE II) and the data informs the district random effects, the scaling factors are all very close to $\lambda = 1$. For PLACE I districts, district-level venue counts are still reasonably close to the truth, although the total number of venues is consistently overestimated in Mangochi and slightly underestimated in Blantyre and Lilongwe. One possible explanation is that there are large district random effects that are not explained by the covariates. The random effect could be related to social, environmental, or political factors that the covariates cannot capture. The difference could also be a result of different efforts in identifying venues in the data collection stage. In particular, Mangochi is the only district where the target number of community informants was not met, potentially leading to an underestimation of the number of venues in the PLACE report. Overall, the model reasonably estimates the total number of venues in each district, even when venue count data is not used, as is the case for PLACE I districts.

\begin{figure}[!tb]
\centering
\begin{subfigure}{.28\textwidth}
  \centering
  \includegraphics[width=1\linewidth]{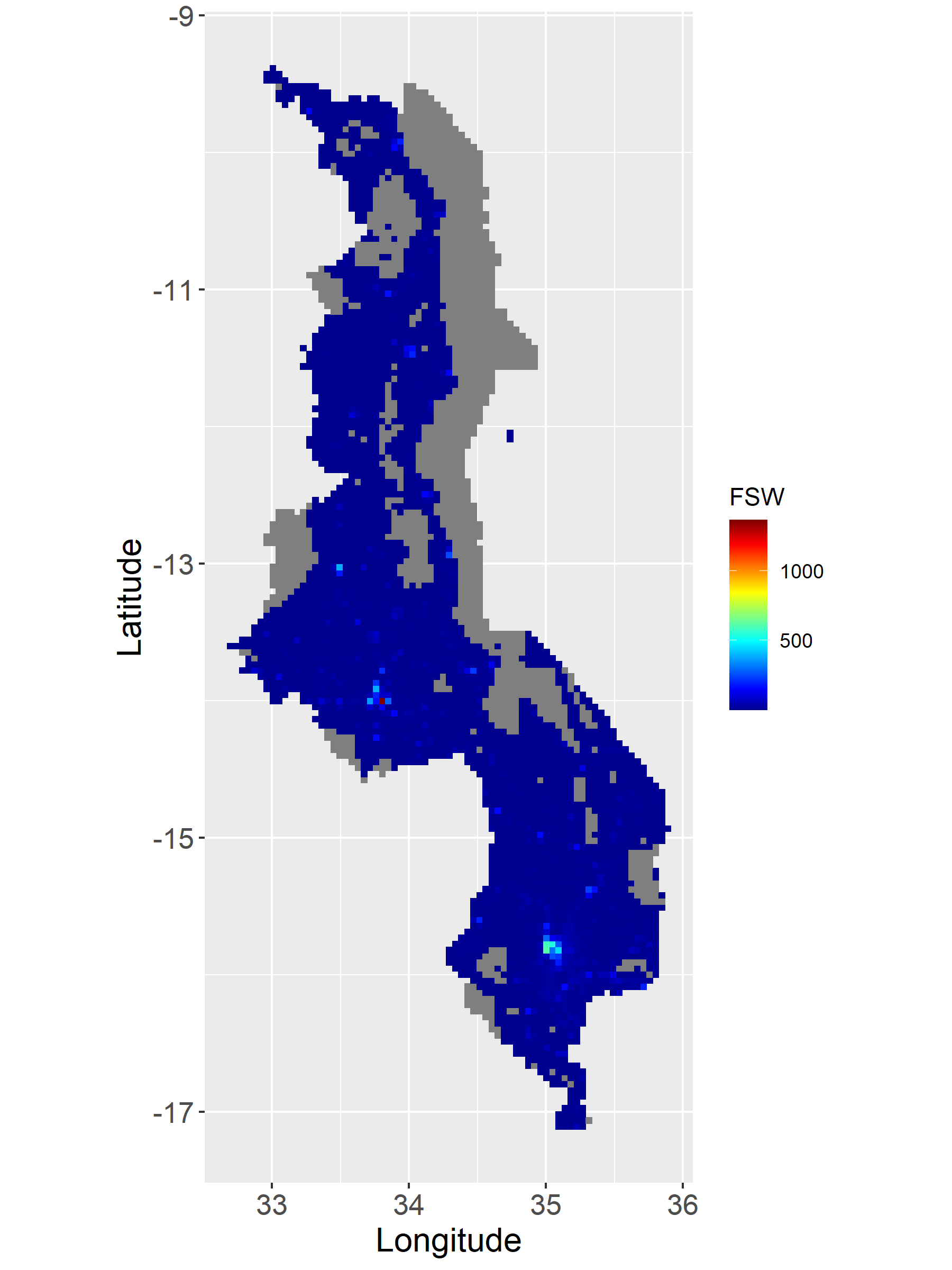}
  \caption{}
  \label{fig:map}
\end{subfigure}%
\begin{subfigure}{.28\textwidth}
  \centering
  \includegraphics[width=1\linewidth]{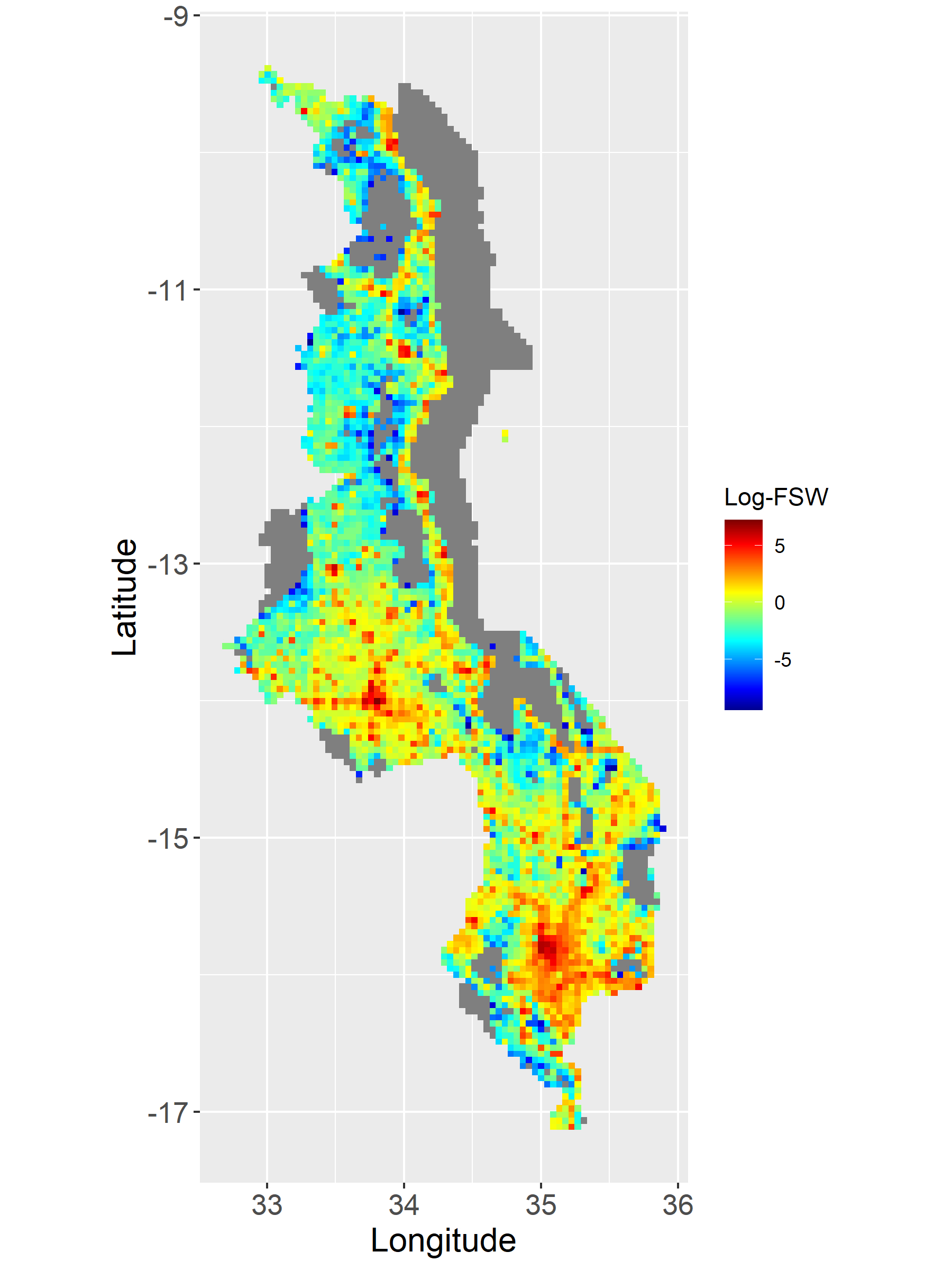}
  \caption{}
  \label{fig:logmap}
\end{subfigure}
\begin{subfigure}{.28\textwidth}
  \centering
  \includegraphics[width=1\linewidth]{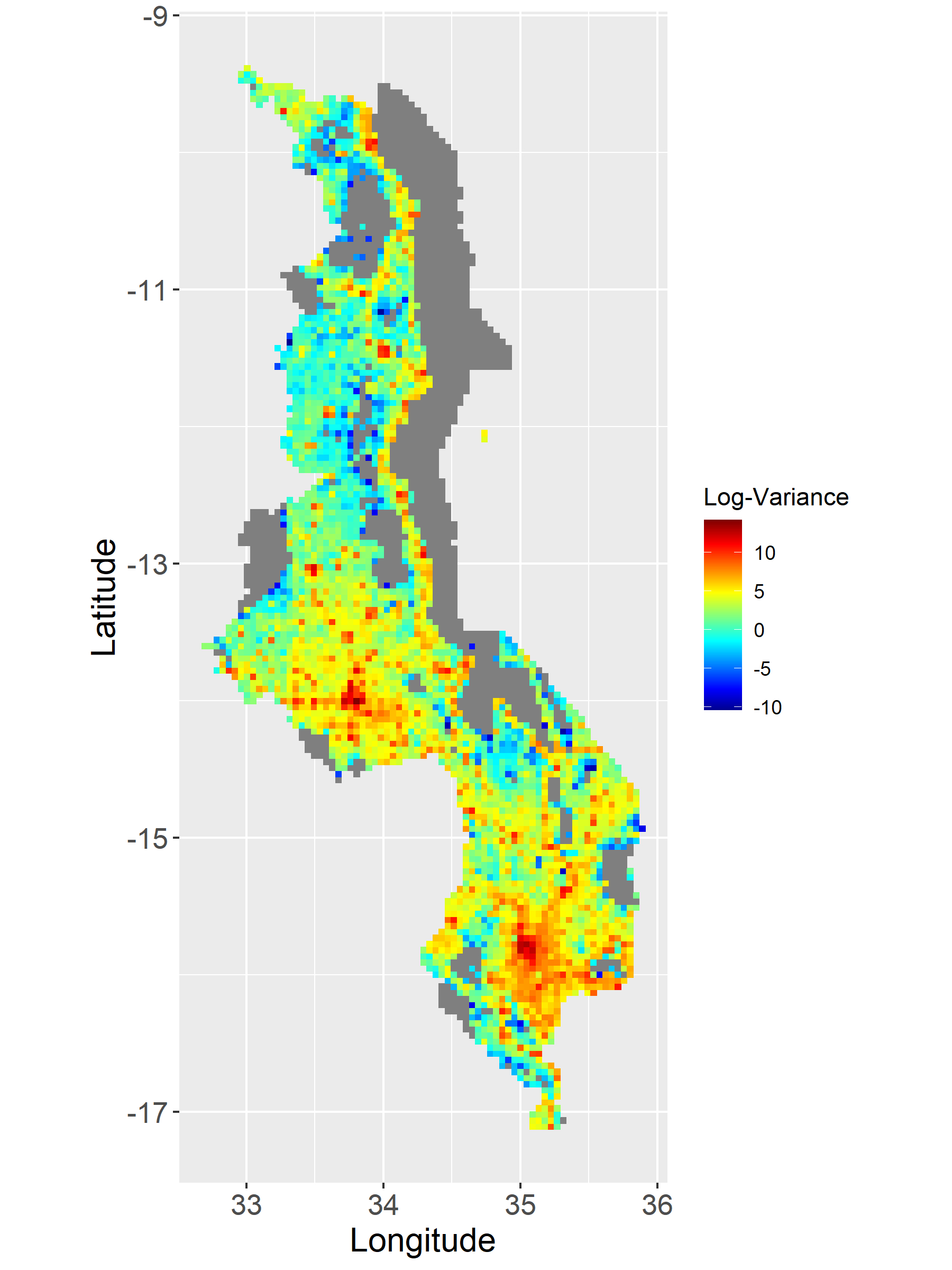}
  \caption{}
  \label{fig:varmap}
\end{subfigure}
\caption{Maps of FSW counts in each cell in Malawi (a) and on the log-scale (b). The log of the variance estimates for FSW counts are shown in (c). Cells with estimated zero FSW are shown as gray.}
\label{fig:fsw_map}
\end{figure}

For privacy reasons, we omit a cell-level mapping of the FSW in Malawi at the resolution used in the model fitting. Instead, we aggregate to much larger cells, approximately $4.8 \times 4.8$-km. Maps of predicted FSW size, FSW size on the log-scale, and the log of the variance estimates of FSW size are shown in Figure \ref{fig:fsw_map}.

\section{Conclusion}
\label{sec:Discussion}
In this paper, we have introduced a fully model-based approach for modeling marked presence-only data. The Bayesian implementation allows us to easily incorporate other data-related issues aside from presence-only, like spatial dependency, while also seamlessly producing credible intervals. The zero-inflated model can, however, be fit using frequentist methods, although confidence intervals and other considerations may be more difficult to calculate.

The results in this paper are critical for both informing broad country and district level targets set by UNAIDS and others as well as local community based outreach programs. National organizations use the larger scale FSW size estimates to properly fund programs to implement HIV prevention and treatment services. The cell-level FSW size estimates are useful for implementing community-based programs like short 30-minute session to teach FSW to negotiate condom use \citep{kerrigan2015community}. Furthermore, fine-scale mapping can further improve the distribution of HIV services. Access to these services is often unequal, resulting in immediate access for some but long wait times, up to years, for others \citep{UNAIDS2021inequality}.

It is important to note that FSW may also work at multiple venues and thus be counted at multiple venues, and FSW who do not work at venues are not counted. Because of this, our goal was not to map the location of all FSW in Malawi, but to map the FSW who attend venues. It is realistic that one may encounter the same FSW at two different venues on two different days due to the mobility. The Malawi PLACE study minimized this impact by asking for the number of FSW who visit the site between 11 p.m. and 2 a.m. on Saturday night. By modeling this data directly, our results also reflect this possibility. However, our analysis does not address FSW net migration across multiple years. FSW do not remain in one location, and major changes can take place due to key events, like the introduction of new policy measures which affect the operation of venues. Thus, our analysis and results provide a reliable map for 2016 and 2017, the time frame of the PLACE study, but do not accurately portray the uncertainty in estimates for other years.

The sampling probabilities $\pi_{d[i]}$ could be refined if a more sophisticated sampling process was used and recorded. Our model can also solve the problem of nonuniform sampling in presence-only studies. Consider a typical presence-only data set where observations come from respondent reports. In these cases, more frequently traveled areas generate more responses. If some measure of ``popularity'' can be created for each cell, then our ZINB model naturally accounts for the varying levels of exposure across the domain. This approach is more natural and theoretically justified given the assumed model than other presence-only approaches, especially those that require choosing pseudo-absences in complicated ways.

The model could also be extended to model venue size and venue count jointly. However, there does appear to be very little correlation between venue size and venue count after accounting for the covariates and this would greatly increase the computation time. Another question that arises is whether interviewing more venues or recording the locations of unvisited venues offers better predictive power. For example, if the survey team can skip an interview at one venue in order to record the location of ten more unvisited venues, would the size of prediction confidence intervals increase or decrease? This is an interesting follow-up question that has real-world application for future sampling procedures.

% \section*{Acknowledgements}
% This research was supported by NIH/NIAID 5-R01-AI136664. The authors thank Dr. Sharon Weir for sharing the PLACE data with us and for helpful discussions. Dr. Weir was vital in helping us understand the collection process of the data as well as informing us about the nature of the HIV epidemic in Malawi. 

% Computations for this research were performed on the Pennsylvania State University’s Institute for Computational and Data Sciences’ Roar supercomputer.

\FloatBarrier

\bibliographystyle{apalike}
\bibliography{biblio.bib}

\newpage
\appendix
\appendixpage

\section{Proofs}

\subsection{Proof of Proposition 1}
\begin{proof}
    Let $Y$ and $X$ be as described in Proposition 1. Then the pmf of $Y$, the thinned process, is given by,
    \begin{align*}
        P(Y = y) &= \sum_{x = 0}^\infty P(Y = y | X = x) P(X = x) \\
            &= \sum_{x = y}^\infty \binom{x}{y} \pi^y (1 - \pi)^{x - y} \binom{x + \phi - 1}{x}\left(\frac{\mu}{\mu  + \phi} \right)^{x} \left(\frac{\phi}{\mu + \phi} \right)^\phi \\
            &= \left(\frac{\pi}{1 - \pi}\right)^y \left(\frac{\phi}{\mu + \phi}\right)^\phi \frac{1}{y! (\phi - 1)!} \sum_{x = y}^\infty \frac{(x + \phi - 1)!}{(x - y)!} \left(\frac{(1 - \pi)\mu}{\mu + \phi} \right)^{x} \\
            &= \left(\frac{\pi}{1 - \pi}\right)^y \left(\frac{\phi}{\mu + \phi}\right)^\phi \frac{1}{y! (\phi - 1)!} (y + \phi - 1)! \left(\frac{\mu(1-\pi)}{\mu + \phi}\right)^y \left(\frac{\mu + \phi}{\mu \pi + \phi}\right)^{y + \phi} \\
            &= \binom{y + \phi - 1}{y}\left(\frac{\pi \mu}{\pi \mu  + \phi} \right)^{y} \left(\frac{\phi}{\pi \mu + \phi} \right)^{\phi},
    \end{align*}
    which is exactly the negative binomial pdf with mean $\pi \mu$ and dispersion $\phi$.
\end{proof}

\subsection{Proof of Proposition 2}
\begin{proof}
    Let $\tilde{Y}$, $Z$, $X$, $X^*$, and $Y$ be as described in Proposition 2. First, let us construct a zero-inflated distribution with the $\pi$-thinning of $X$, $Y^* = (1 - Z)X^* = (1-Z) \sum_{i=1}^X I_{y^*,i}$. Second, consider the $\pi$-thinning of the original zero-inflated distribution, $Y = \sum_{i=1}^{\tilde{Y}} I_{y,i} = \sum_{i=1}^{(1-Z)X} I_{y,i}$. Conditioning on $Z$ for both distributions, we find
    \begin{equation*}
        \begin{split}
            P(Y^* = y | Z = 0) &= 1 - p \\
            P(Y^* = y | Z = 0) &= P\left(\sum_{i=1}^X I_{y^*,i} = y\right)p
        \end{split}
    \end{equation*}
    and
    \begin{equation*}
        \begin{split}
            P(Y = y | Z = 0) &= 1 - p \\
            P(Y = y | Z = 0) &= P\left(\sum_{i=1}^X I_{y,i} = y\right)p
        \end{split}
    \end{equation*}
    
    Since $I_{y^*,i} \overset{D}{=} I_{y,i}$, it follows that $Y^* \overset{D}{=} Y$.
\end{proof}

\subsection{Proof of Proposition 3}
\begin{proof}
    Let $\tilde{Y}$, $Y$, and $I_k$ be as defined in Theorem 3. From Proposition 2, $Y$ is also a zero-inflated random variable, $Y = (1-Z)X^*$, where $X^* = \sum_{i=1}^X I_{x,i}$, $I_{x,i} \sim Bern(\pi)$. From Proposition 1, $X^* \sim NegBin(\text{mean} = \pi \mu, \text{dispersion} = \phi)$. Thus, $Y$ is a zero-inflated negative binomial negative binomial, with negative binomial mean $\pi \mu$ and dispersion $\phi$.
\end{proof}

\clearpage
\newpage
\section{Additional Diagnostics}

\begin{table}[!hb]
\centering
\caption{Posterior predictive p-values for the complete districts.}
\begin{tabular}{lcc}
\hline \hline
\textbf{Test Statistics}  & \textbf{Venue Count} & \textbf{Venue Size}  \\ \hline
Positive Mean & 0.537  & 0.546 \\
Positive Standard Deviation & 0.560  & 0.514 \\
Over-dispersion index & 0.566   & 0.508\\
Maximum & 0.496 &  0.584 \\
Proportion of Zeros & 0.482  & 0.483 \\ \hline
\end{tabular}
\label{tab:complete_ppp}
\end{table}

\bigskip

\begin{figure}[!hb]
\centering
\begin{subfigure}{.24\textwidth}
  \centering
  \includegraphics[width=1\linewidth]{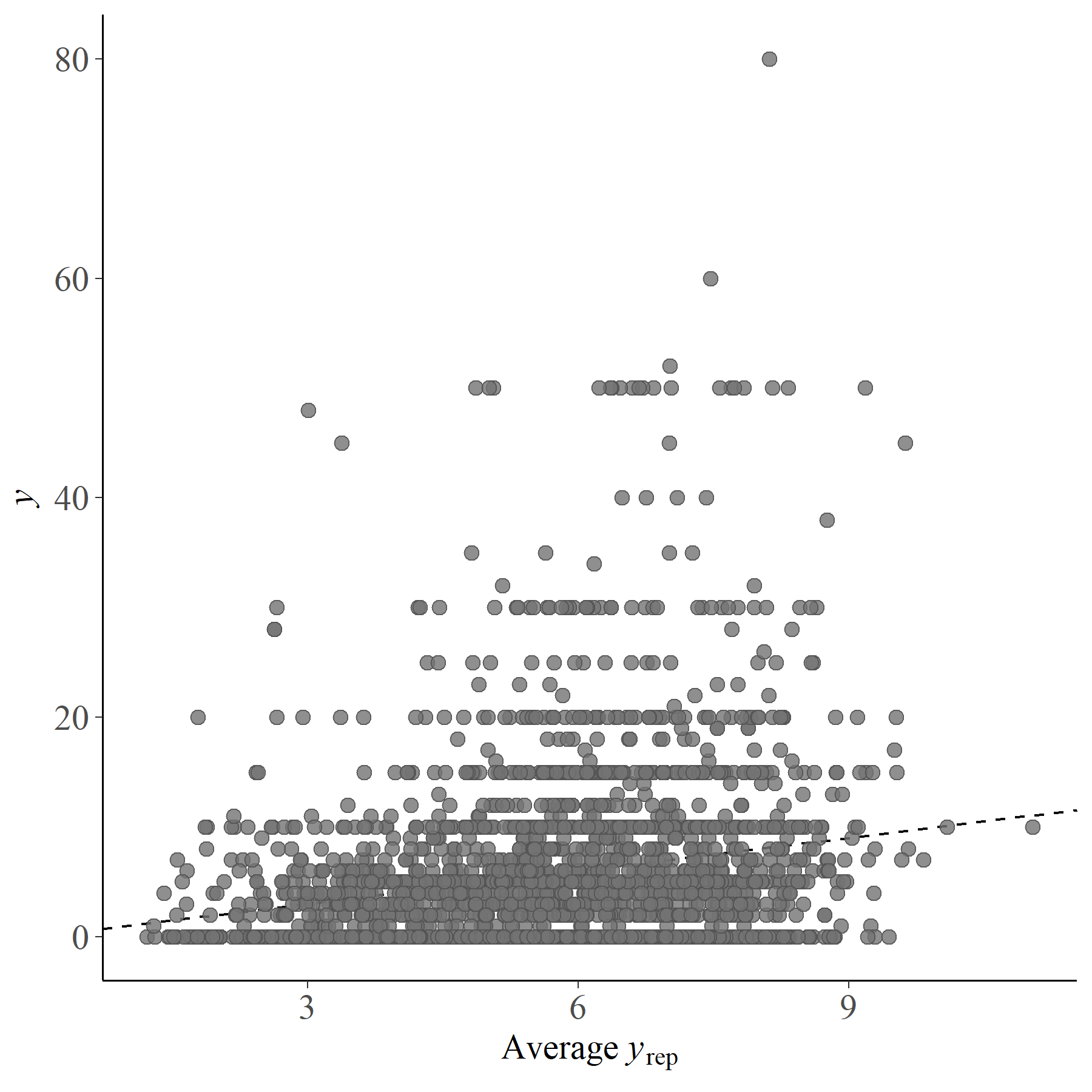}
  \caption{}
  \label{fig:diaga}
\end{subfigure}%
\begin{subfigure}{.24\textwidth}
  \centering
  \includegraphics[width=1\linewidth]{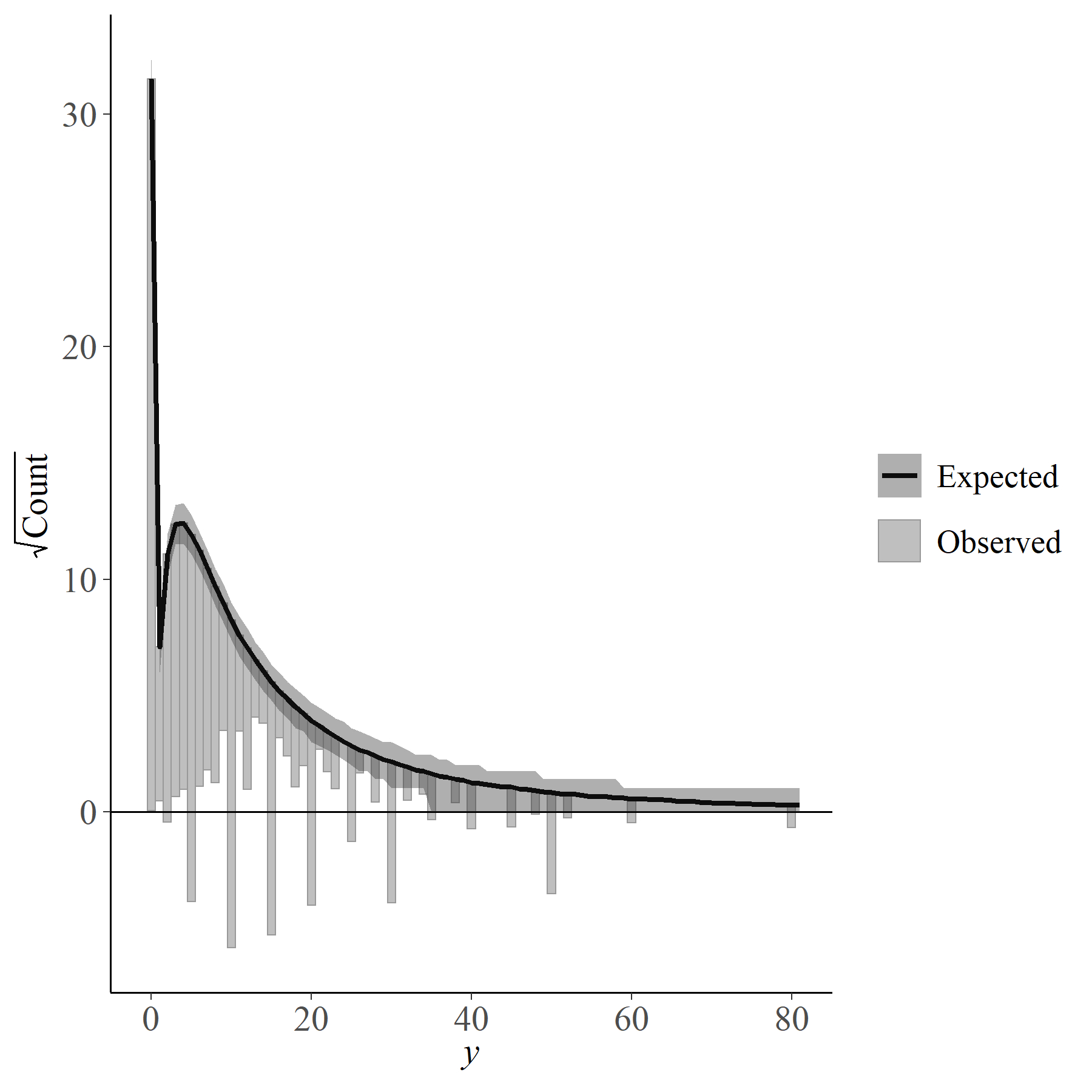}
  \caption{}
  \label{fig:diagb}
\end{subfigure}
\begin{subfigure}{.24\textwidth}
  \centering
  \includegraphics[width=1\linewidth]{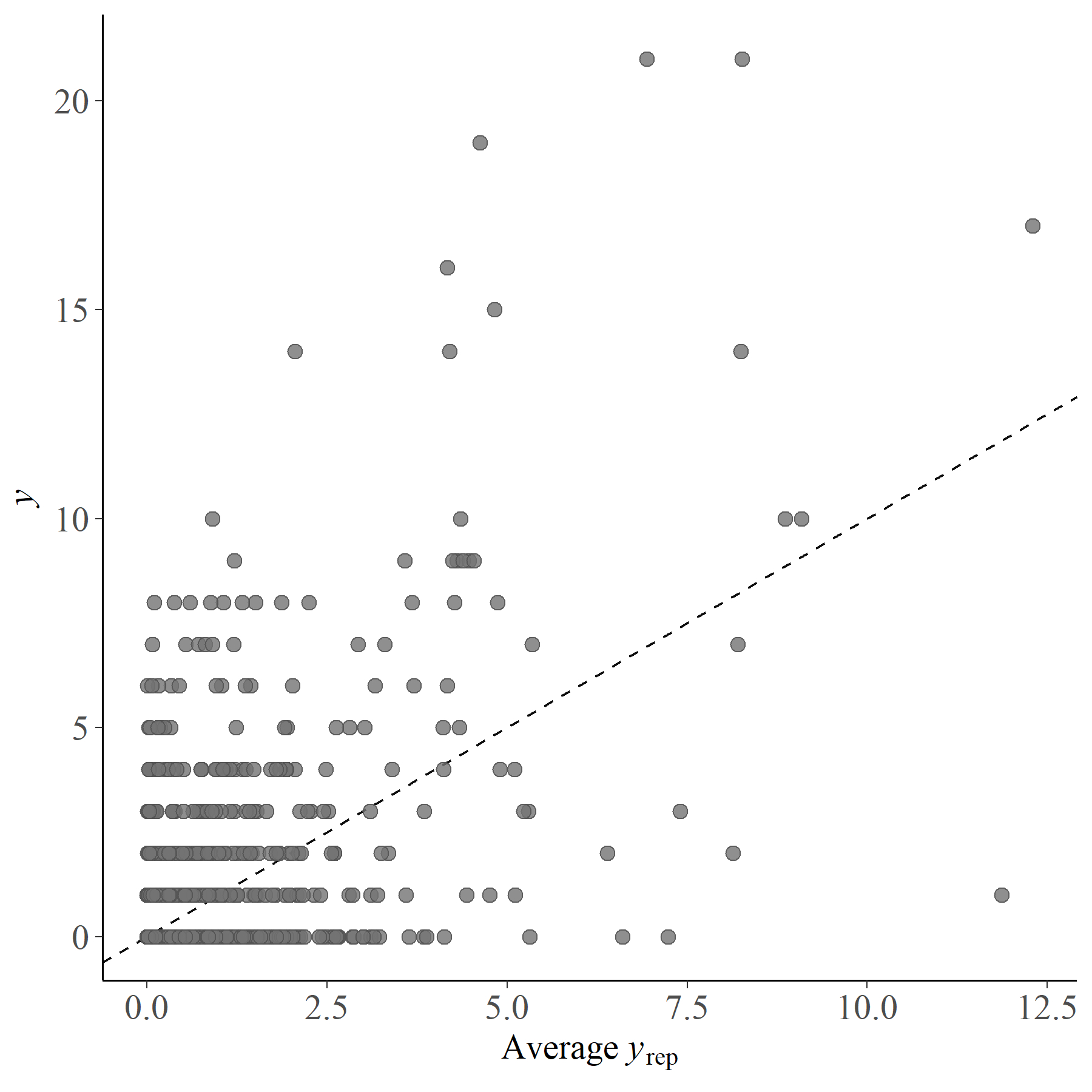}
  \caption{}
  \label{fig:diagc}
\end{subfigure}
\begin{subfigure}{.24\textwidth}
  \centering
  \includegraphics[width=1\linewidth]{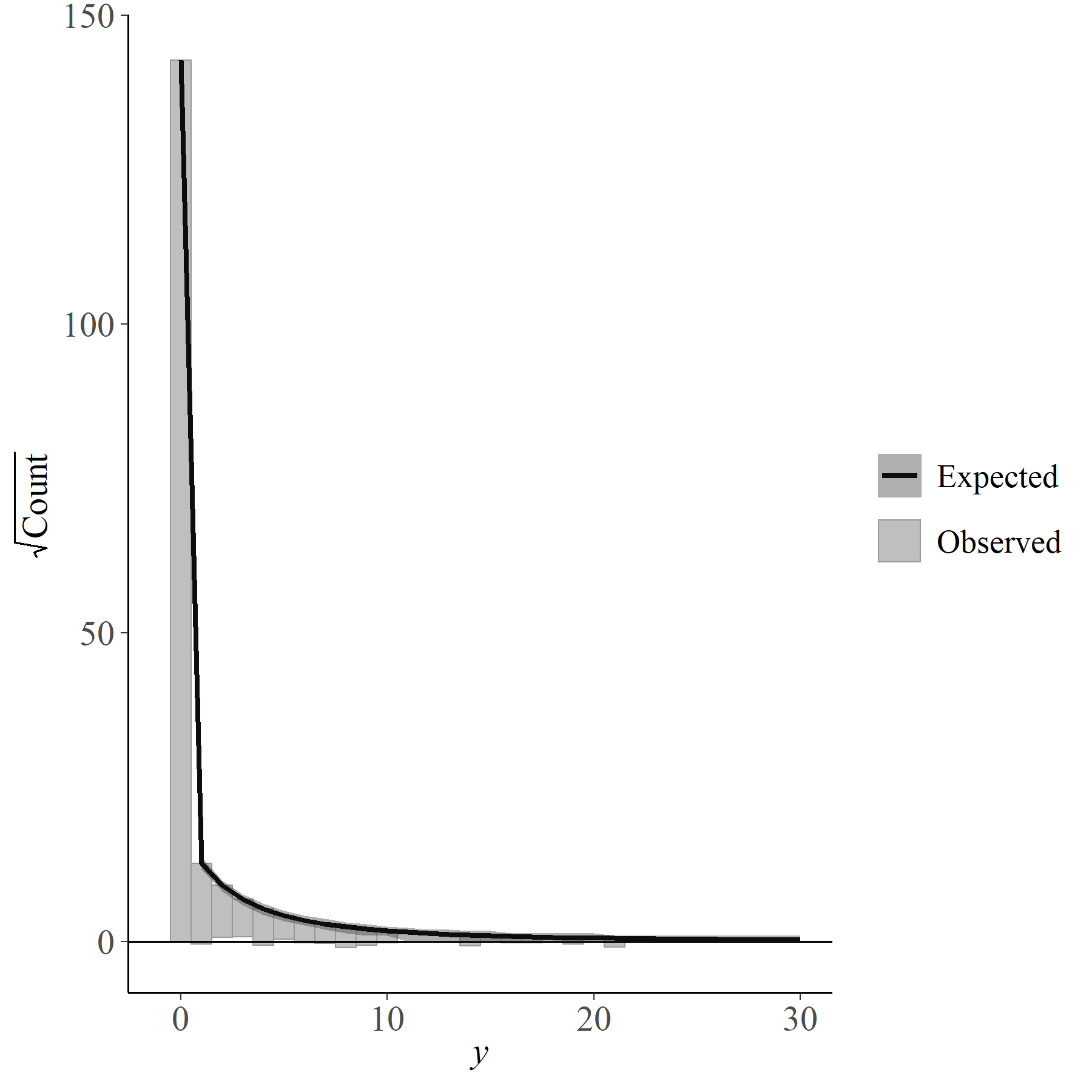}
  \caption{}
  \label{fig:daigd}
\end{subfigure}
\caption{Diagnostic plots for venue size and count. Subplots (a) and (b) concern the venue size while subplots (c) and (d) concern the venue count. (a) and (c) are scatter plots of average predicted values against the observed predicted counts. (b) and (c) are hanging rootograms and plot the square roots of the counts against the response variable. The observed counts are shown as light gray bars while the expected counts and corresponding uncertainty are shown as a black line with dark gray shadow.}
\label{fig:diags}
\end{figure}

\end{document}